%% file: main.tex
\documentclass[a4paper,11pt]{article}
\pdfoutput=1 

\usepackage{jcappub} 

\usepackage[T1]{fontenc} 

\usepackage{amsmath}
\usepackage{amssymb}
\usepackage{subcaption} 
\usepackage{graphicx}
\usepackage[all]{nowidow}
\usepackage[utf8]{inputenc}
\usepackage{multicol}

\definecolor{ShamrockGreen}{rgb}{0.0, 0.62, 0.38}

\title{Constraining axion-like particles using the white dwarf initial-final mass relation}

\author{Matthew J. Dolan,}
\author[1]{Frederick J. Hiskens,\note{Corresponding author}}
\author{and Raymond R. Volkas}
\affiliation{ARC Centre of Excellence for Dark Matter Particle Physics, School of Physics, The University of Melbourne, Victoria 3010, Australia}

\emailAdd{dolan@unimelb.edu.au}
\emailAdd{fhiskens@student.unimelb.edu.au}
\emailAdd{raymondv@unimelb.edu.au}

\arxivnumber{2102.00379}

\abstract{Axion-like particles (ALPs), a class of pseudoscalars common to many extensions of the Standard Model, have the capacity to drain energy from the interiors of stars. Consequently, stellar evolution can be used to derive many constraints on ALPs. We study the influence that keV-MeV scale ALPs which interact exclusively with photons can exert on the helium-burning shells of asymptotic giant branch stars, the late-life evolutionary phase of stars with initial masses less than $8M_{\odot}$. We establish the sensitivity of the final stellar mass to such energy-loss for ALPs with masses currently permitted by stellar evolution bounds. A semi-empirical constraint on the white dwarf initial-final mass relation (IFMR) derived from observation of double white dwarf binaries is then used to exclude part of a currently unconstrained region of ALP parameter space, the cosmological triangle. The derived constraint relaxes when the ALP decay length becomes shorter than the width of the helium-burning shell. Other potential sources for stellar constraints on  ALPs are also discussed.}

\begin{document}
\maketitle
\flushbottom

\input{Introduction/Introduction}
\input{Section2/Sec2}
\input{Section3/Sec3}
\input{Section4/Sec4}
\input{Section5/Sec5}
\input{Section6/Sec6}

\appendix
\input{AppA/AppA}
\input{AppB/AppB}
\input{AppC/AppC}

\acknowledgments
We would like to thank Katie Auchettl for her invaluable insight and assistance, as well as Jeff Andrews, Pierluca Carenza and Felix Kahlhoefer for helpful comments and discussion. This work was supported in part by the Australian Research Council and the Australian Government Research Training Program Scholarship initiative.

\bibliographystyle{JHEP}
\bibliography{bibliography}
\end{document}

%% file: Introduction/Introduction.tex
\section{Introduction}
\thispagestyle{plain}
\begin{figure}[b]
    \centering
    \includegraphics{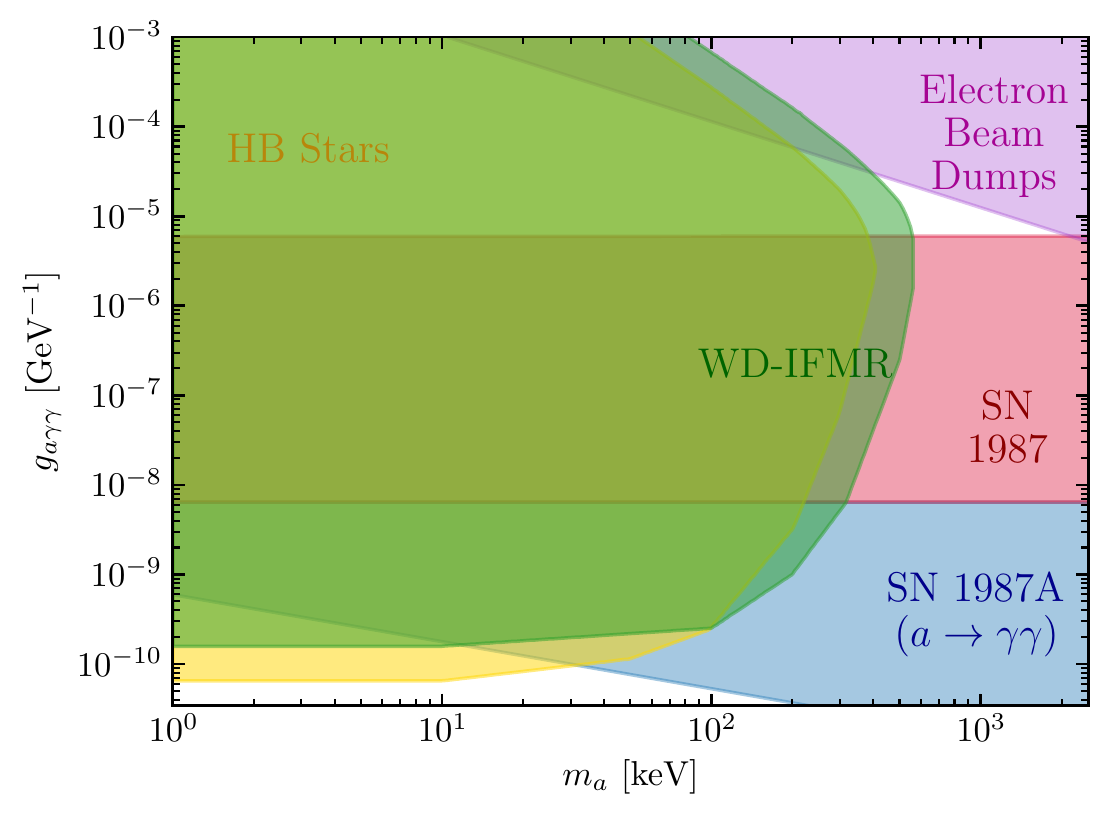}
    \caption{Constraints on ALP mass $m_a$ and coupling strength to photons $g_{a\gamma\gamma}$ in the keV-MeV mass range. Individual bounds are referenced in the text. These are shown at 95\% confidence level. The constraint derived in this work is labelled 'WD-IFMR'.}
    \label{fig: ALP_param_space}  
\end{figure}

Axion-like particles (ALPs) are light, weakly interacting pseudoscalars which feature in many extensions of the Standard Model (SM) of particle physics. They arise as pseudo-Nambu Goldstone bosons (pNGBs) of spontaneously broken symmetries in, for example, the Peccei-Quinn solution of the strong CP problem \cite{PQ1, Peccei:1977ur, Weinberg-40.223, Wilczek:1977pj}, compactification scenarios in string theory \cite{Svrcek:2006yi, Arvanitaki:2009fg, Cicoli:2012sz} and in models of electroweak relaxation \cite{Graham:2015cka}.

The properties of specific ALPs, such as their masses and coupling strengths to SM particles, are model-dependent, which has sparked investigations of their influence in a wide phenomenological range. Light ALPs with masses below the MeV scale impact astrophysical and cosmological phenomena \cite{Cadamuro:2011fd}, such as Big Bang Nucleosynthesis (BBN) \cite{Updated_BBN}, the Cosmic Microwave Background (CMB) and stellar evolution \cite{Raffelt-Bounds-on-light, RAFFELT1982323, Raffelt:1996wa, Ayala:2014pea, Aoyama:2015asa, Carenza:2020zil, Friedland:2012hj, Dominguez, Dominguez:2017mia}. As pNGBs, they can be naturally light and weakly interacting, which makes them ideal candidates for cold dark matter (DM) \cite{Preskill:1982cy, Abbott:1982af, Dine:1982ah, Arias:2012az}.

ALPs in the MeV to GeV range, however, are generally too massive to significantly influence cosmology and astrophysics, yet are relevant in aspects of particle physics. It has been suggested that ALPs may contribute to the anomalous muon magnetic moment \cite{Chang:2000ii, Bauer:2017nlg, Marciano:2016yhf}, or act as a portal between the dark sector and SM particles \cite{Nomura:2008ru}. Theoretical and experimental interest in ALPs has risen significantly with the suggestion that DM ALPs can explain the recent $3.5\sigma$ excess in electron-recoil measured at the XENON1T experiment \cite{Athron:2020maw, Takahashi:2020uio, Takahashi:2020bpq}.

In this work our attention is limited to ALPs which couple exclusively to photons via the interaction
\begin{equation}
    \label{eq: ALP-Photon interaction}
    \mathcal{L}_a=-\frac{g_{a\gamma\gamma}}{4}F_{\mu\nu}\Tilde{F}^{\mu\nu}a,
\end{equation}
where $g_{a\gamma\gamma}$ is the ALP-photon coupling strength, $a$ is the ALP-field, $F_{\mu\nu}$ is the electromagnetic field-strength tensor and $\tilde{F}^{\mu\nu}$ its dual. Specifically, we consider such ALPs with masses $m_a$ in the keV-MeV range, the $m_a$-$g_{a\gamma\gamma}$ parameter space of which is shown in Figure \ref{fig: ALP_param_space}. Constraints for this region arise from stellar evolution \cite{Carenza:2020zil}, results at electron beam dumps, in particular the SLAC E137 experiment \cite{Bjorken87, Dolan:2017osp} and the neutrino signal associated with the cooling of SN1987A \cite{Lucente:2020whw} as well as the visible signal associated with a subsequent decay of ALPs into photons \cite{Jaeckel:2017tud}.

These bounds fail to exclude a small triangular region at $g_{a\gamma\gamma}\sim10^{-5}\ \mathrm{GeV}^{-1}$ and $m_a\sim1\ \mathrm{MeV}$, referred to as the \textit{cosmological triangle}. While constraints derived from BBN exclude the cosmological triangle \cite{Cadamuro:2011fd, Updated_BBN}, these relax significantly in certain scenarios of non-standard cosmology \cite{Dolan:2017osp, Depta:2020wmr}. Furthermore, as several approaching experiments will have the capacity to directly probe the cosmological triangle \cite{Dolan:2017osp, Brdar:2020dpr}, it is timely to investigate  phenomenological implications of ALPs within this region. In this paper we revisit the effects ALPs can exert on stellar evolution.

Over nearly four decades, stellar evolution has been frequently deployed to constrain ALPs via the so-called \textit{energy-loss argument}. The general mechanism by which this operates is as follows, though a more detailed account of its effects on stellar structure can be found in \cite{Raffelt:1996wa} and specific production process will be discussed in Section \ref{sec: Section 2}. If sufficiently light and weakly interacting, ALPs produced in stellar interiors can freely escape the star and act as a local energy-sink in that region. This merely results in cooling of the stellar plasma if the zone in which production occurred hosts no nuclear activity. If, on the other hand, this region is undergoing nuclear burning, the star accounts for the energy deficit by contracting and heating, which drives the intensity of fusion upwards. As a result both the rate of consumption of nuclear fuel and the energy-loss rate associated with ALP-production increase and a positive feedback mechanism is defined, which ultimately accelerates the progression of the entire evolutionary phase. For sufficiently strongly interacting ALPs, this effect introduces a contradiction between theory and observation, which leads to a constraint.

The most stringent stellar energy-loss constraints on ALPs interacting with photons alone have been derived from observation of horizontal branch (HB) stars \cite{Raffelt-Bounds-on-light, Raffelt:1996wa, Cadamuro:2011fd, Ayala:2014pea, Carenza:2020zil}. Specifically, population studies in globular clusters (large gravitationally bound collections of old, metal-poor stars) place limits on the helium-burning lifetime $\tau_{\mathrm{He}}$ of $0.8M_{\odot}$ HB stars. If, for a given choice of $m_a$ and $g_{a\gamma\gamma}$, ALP energy-loss causes $\tau_{\mathrm{He}}$ to fall below the observed lower bound, it can be excluded. This is exactly the basis of the HB star constraint in Figure \ref{fig: ALP_param_space}. Notably this relaxes as $m_a$ increases beyond the HB star core temperature of 10 keV, owing to the Boltzmann suppression of ALP-production.

However, ALPs are known to influence many more aspects of stellar evolution than just HB stars. ALPs as massive as 100 MeV can drain energy from the cores of supernovae (SN). The magnitude of novel energy-loss, however, is constrained by the SN1987A neutrino signal, the standard astrophysical source of supernova cooling. The SN1987A ALP constraint shown in Figure \ref{fig: ALP_param_space} was recently computed using a state-of-the-art SN model in \cite{Lucente:2020whw}. Note that it does not extend to arbitrarily high values of $g_{a\gamma\gamma}$. Instead, ALPs become trapped within the supernova core and contribute towards energy transfer. The constraint relaxes when this ALP energy transfer falls below that of neutrinos. This defines the lower $g_{a\gamma\gamma}$ boundary of the cosmological triangle and motivates the search for complementary constraints.

Beyond these, it is also known that ALP production can alter chemical abundances during nucleosynthesis in core collapse supernova progenitors \cite{Aoyama:2015asa}, prevent the blue loop evolutionary phase from occurring in intermediate mass helium-burning stars \cite{Friedland:2012hj}, and cause significant structural changes in late-life intermediate mass stars on the asymptotic giant branch (AGB) \cite{Dominguez, Dominguez:2017mia}. These investigations, however, have been restricted to low-mass ALPs ($m_a\lesssim10$ keV) and their potential for constraining MeV-scale ALPs has never been assessed.

We rectify this by exploring the influence of keV-MeV mass ALPs on stars on the asymptotic giant branch, the late-life evolutionary phase of stars with masses $\lesssim8M_{\odot}$. It has long been recognised that the production of low-mass ALPs within the helium-burning (He-B) shell of such stars can greatly affect their final mass $M_\mathrm{f}$ \cite{Dominguez}. Though this tendency is likely to have observable consequences for white dwarfs (WDs) and core collapse supernova (CCSN) progenitors, it has not led to the construction of a robust ALP constraint. The He-B shells of intermediate mass AGB stars, however, are typically hotter than the cores of HB stars in globular clusters, making them an enticing prospect for further constraining the cosmological triangle.

We simulate stellar evolution with and without energy-loss to massive ALPs to establish the sensitivity of $M_{\mathrm{f}}$ in a region of the ALP-plane unrestricted by the HB star bound, using an edited version of the open source, 1-D stellar evolution code \textit{Modules for Experiments in Stellar Astrophysics} (\texttt{MESA})  \cite{MESA1, MESA2, MESA3, MESA4, MESA5}. \texttt{MESA} is a suite of modules containing up-to-date astrophysics such as opacity tables, nuclear reaction rates and equations-of-state. Its stellar evolution module \texttt{MESAstar} has shown remarkable versatility at modelling stars over a wide range of initial masses. It has been widely tested and compared to astrophysical observation and other stellar evolution codes.

A constraint on ALPs based on reducing $M_{\mathrm{f}}$ is then established by using the white dwarf initial-final mass relation (IFMR). The IFMR maps the initial mass with which a star forms $M_{\mathrm{init}}$ to the final mass of the white dwarf into which it ultimately evolves. It is used in age and distance determination in globular clusters and informs our understanding of supernovae rates \cite{Greggio-TypeIa}, galactic chemical evolution and the field white dwarf population. Numerous constraints on the IFMR exist \cite{Weidemann2000, Kalirai_2008, Williams2009, Andrews, cummings_2016, El-Badry, Cummings_2018} and have previously been used to restrict stellar physics on the AGB \cite{Cummings_2019}. We redeploy one of these constraints, derived from wide double white dwarf binaries, to produce a robust bound on ALPs derived from AGB stars.

This paper has the following structure. In Section \ref{sec: Section 2} we describe the mechanisms of ALP photo-production in a stellar plasma. We then use the results of our simulations in \texttt{MESA} to explore the impact of MeV scale ALPs on the AGB in Section \ref{sec: Sec 3}. In Section \ref{sec: Section 4} we construct our constraint, account for the possibility of ALP-decay and detail some of the systematic uncertainties affecting our analysis. Other possible observational constraints are mentioned in Section \ref{sec: Section 5} before we summarise and conclude in Section \ref{sec: Section 6}. We describe our treatment of ALP energy-loss in MESA, as well as our adopted input physics in Appendices \ref{Sec: AppA} and \ref{sec: AppB} respectively. Further discussion of the systematic uncertainties relevant to this work is included in Appendix \ref{sec: AppC} while Appendix \ref{sec: AppD} includes an alternative probabilistic approach to deriving a constraint.

%% file: Section2/Sec2.tex
\section{Astrophysical ALP Production}
\label{sec: Section 2}

The impact of ALPs on stellar structure varies with their lifetime $\tau_a=\Gamma_a^{-1}$ and production cross-section. The interaction in Equation \ref{eq: ALP-Photon interaction} leads to the decay width
\begin{equation}
    \label{eq: ALP-decay width}
    \Gamma_a=\frac{g_{a\gamma\gamma}^2m_a^3}{64\pi}.
\end{equation}
For sufficiently small values of $m_a$ and $g_{a\gamma\gamma}$, this lifetime is large and ALPs freely escape the stellar interior and contribute to energy-loss within the star. When the ALP mass and photon-coupling grow large, however, ALPs decay within the star and contribute towards radiative energy-transfer.

\subsection{Energy-loss to ALP-production}

The production of freely-escaping ALPs affects stellar structure by reducing the local energy-gain rate per unit mass $\epsilon$. The magnitude of the ALP energy-loss rate $\epsilon_a$ is given as a sum over the relevant ALP-production processes. For ALPs in the keV-MeV range which couple only to photons two such mechanisms of significance exist, \textit{Primakoff production} and \textit{photon coalescence} (or fusion).

ALP-Primakoff production refers to the conversion of a photon into an ALP in the presence of an external electromagnetic field \cite{Dicus-PhysRevD.18.1829, Raffelt:1996wa}. In a stellar interior this is facilitated by the Coulomb field of the constituent charged particles in the plasma. The transition rate of a photon with momentum $\Vec{k}$ and energy $\omega$ into an ALP of momentum $\Vec{p}$ is \cite{DiLella:2000dn} 
\begin{equation}
\begin{split}
    \Gamma_{\gamma\rightarrow a}^P=\frac{g_{a\gamma\gamma}^2Tk_s^2}{32\pi}\frac{k}{\omega}
    \Bigg(\frac{((k+p)^2+k_s^2)((k-p)^2+k_s^2)}{4kpk_s^2}&\ln\bigg(\frac{(k+p)^2+k_s^2}{(k-p)^2+k_s^2}\bigg)\\-\frac{(k^2-p^2)^2}{4kpk_s^2}\ln\bigg(\frac{(k+p)^2}{(k-p)^2}\bigg)-1
    \Bigg),
\end{split}
\label{eq: Primakoff Transition Rate}
\end{equation}
where $k=\lvert\Vec{k}\rvert$, $p=\lvert\Vec{p}\rvert$ and $T$ is the temperature of the stellar plasma. Importantly, the Primakoff transition rate is subject to plasma screening effects, the scale of which is set by the Debye-H\"uckel wave number
\begin{equation}
    k_s^2=\frac{4\pi\alpha}{T}\frac{\rho}{m_u}\bigg(Y_e + \sum_jZ_j^2Y_j\bigg),
\end{equation}
where $\rho$ is the local mass density, $m_u$ is the atomic mass unit, $Y_e$ is the number of electrons per baryon and $Y_j$ is the number per baryon of nuclear species with charge $Z_j$.

In the stellar plasma, the effective photon mass is given by the plasma frequency $\omega_{\mathrm{pl}}\approx4\pi\alpha n_e/m_e$. In all scenarios we consider, however, this is small compared with the average photon energy and is therefore neglected. Furthermore, we assume that the mass of the produced ALP is small compared with that of the charged particle. Consequently, its recoil can be ignored and Equation \ref{eq: Primakoff Transition Rate} simplifies to
\begin{equation}
\begin{split}
    \Gamma_{\gamma\rightarrow a}^P=\frac{g_{a\gamma\gamma}^2Tk_s^2}{32\pi}     \Bigg(\frac{(m_a^2-k_s^2)^2+4\omega^2k_s^2}{4\omega pk_s^2}&\ln\bigg(\frac{(\omega+p)^2+k_s^2}{(\omega-p)^2+k_s^2}\bigg)\\-\frac{m_a^4}{4\omega pk_s^2}\ln\bigg(\frac{(\omega+p)^2}{(\omega-p)^2}\bigg)-1
    \Bigg)
\end{split}
\label{eq: Simplified Primakoff Transition Rate}
\end{equation}
where $p=\sqrt{\omega^2-m_a^2}$.

\begin{figure}[t]
    \centering
    \begin{subfigure}[t]{0.5\textwidth}
        \centering
        \includegraphics{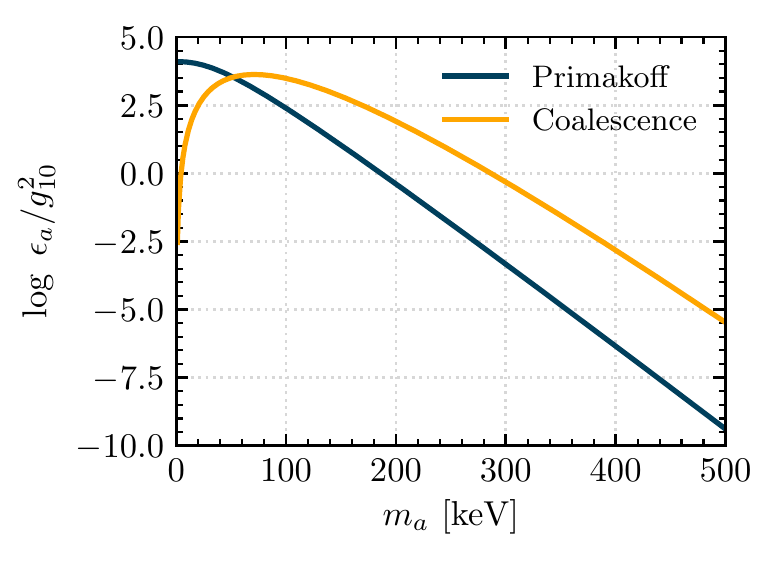}
        \caption{}
        \label{fig: Eps_processcomp}
    \end{subfigure}%
    \begin{subfigure}[t]{0.5\textwidth}
         \centering
         \includegraphics{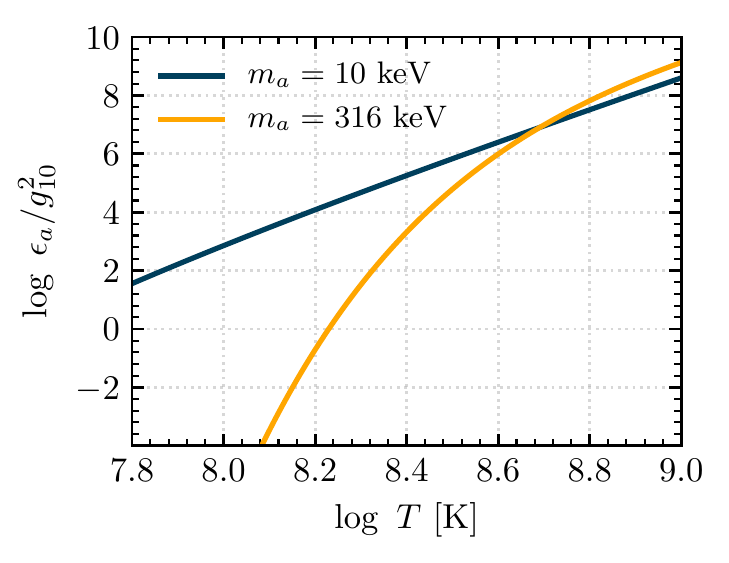}
         \caption{}
         \label{fig: Eps_masscomp}
     \end{subfigure}
    \caption{The magnitude of energy-loss rate per unit mass associated with ALP-production for the Primakoff and Coalescence production mechanisms as a function of ALP mass $m_a$ given conditions within the He-B layer of a $4M_{\odot}$ AGB star (a). A comparison between the total energy-loss rates per unit mass is also shown as a function of temperature given two different ALP masses, 10 keV and 316 keV (b).}
    \label{fig: Fig2}
\end{figure}

The contribution of this process to $\epsilon_a$ can then be computed as \cite{Raffelt-Bounds-on-light, Raffelt:1996wa}
\begin{equation}
    \epsilon_a^P=\frac{2}{\rho}\int\frac{dp\ p^2}{2\pi^2}\Gamma_{\gamma\rightarrow a}\omega f(\omega),
\end{equation}
where the factor of 2 accounts for the photon polarisations and $f(\omega)$ is the thermal photon energy distribution, here given by the Bose-Einstein distribution \cite{DiLella:2000dn}
\begin{equation}
    f(\omega)=\frac{1}{e^{\omega/T}-1}.
\end{equation}
Substituting Equation \ref{eq: Simplified Primakoff Transition Rate} into the expression for $\epsilon_a^P$ then gives
\begin{equation}
    \epsilon_a^P=\frac{g_{a\gamma\gamma}^2T^7}{4\pi\rho}F(\xi^2, \mu^2)
    \label{eq: F function}
\end{equation}
where dimensionless parameters $\mu=m_a/T$ and $\xi=k_s/(2T)$ have been defined. The function $F$ involves an integral over photon phase-space and encompasses the entire $m_a$- and $k_s$-dependence of $\epsilon_a$.

So long as the kinetic threshold $m_a\geq2\omega_{\mathrm{pl}}$ is met, photon coalescence $\gamma \gamma \to a$ can contribute to ALP production in stellar interiors. The production rate due to this process is \cite{DiLella:2000dn}
\begin{equation}
    \frac{dN_a}{d\omega}=\frac{g_{a\gamma\gamma}^2m_a^4}{128\pi^3}\sqrt{\omega^2-m_a^2}e^{-\omega/T},
\end{equation}
which corresponds to an energy-loss rate
\begin{equation}
    \epsilon_{a}^C=\frac{1}{\rho}\int\omega\frac{dN_a}{d\omega}d\omega=\frac{g_{a\gamma\gamma}^2T^7}{4\pi\rho}G(\mu^2).
    \label{eq: G function}
\end{equation}
Here $G(\mu^2)$, like $F(\xi^2, \mu^2)$ above, contains the entire $m_a$-dependence of $\epsilon_a^C$. The total energy-loss rate per unit mass to ALP production is then given by
\begin{equation}
    \epsilon_a=\frac{g_{a\gamma\gamma}^2T^7}{4\pi\rho}(F(\xi^2, \mu^2)+G(\mu^2)).
\end{equation}
Both functions $F(\xi^2, \mu^2)$ and $G(\mu^2)$ contain integrals over photon phase-space which must be evaluated numerically.

The relative importance of these two production mechanisms was comprehensively discussed in \cite{Carenza:2020zil}. Primakoff production was found to dominate energy-loss in HB star cores for low ALP masses ($\approx30$~keV). When ALPs with masses of $80$~keV were considered, however, photon coalescence contributed most significantly towards $\epsilon_a$. As can be seen in Figure \ref{fig: Eps_processcomp}, the same is true when the conditions of the He-B shell of a $4M_{\odot}$ AGB star are adopted ($T\approx16$~keV, $\rho\approx1.4\times10^3$ g cm$^{-3}$, $k_s\approx35$~keV) when normalised by $g_{10}^2\equiv (g_{a\gamma\gamma}/[10^{-10}\ \mathrm{GeV}^{-1}])^2$. Note that these values have been taken from our models.

Both production mechanisms are Boltzmann suppressed for heavy ALPs. Consequently the temperature sensitivity of $\epsilon_a$ is enhanced significantly as $m_a$ increases. This is depicted for 10~keV and 316~keV ALPs in Figure \ref{fig: Eps_masscomp} given the conditions of the HeB shell. Production of the latter, which is dominated by photon fusion, rapidly increases between $\log\ T=8.2$ and $8.6$ as Boltzmann suppression is alleviated. Energy-loss to 10 keV ALPs, however, increases far more gradually over this temperature interval and falls below that of the 316 keV ALPs at high temperatures. It is precisely this temperature dependence which motivates why moderate increases in temperature can elicit strong improvements in existing constraints.

It should be noted, however, that this calculation has assumed a constant Debye-H\"uckel wave number, which is not true in a stellar environment. Consequently, the temperature normalised Debye-H\"uckel wave number $\xi=k_s/(2T)$ decreases, which somewhat inhibits Primakoff ALP production, resulting in total energy-loss for the 10~keV ALP being underestimated. The value of 316~keV has been chosen because $316=10^{2.5}$ and we shall be investigating the impact of ALPs logarithmically spaced in mass.

\subsection{ALP contributions to energy-transport}
Strongly interacting ALPs, which decay before departing the local stellar region, modify stellar structure by contributing an additional term to the radiative opacity of the medium. The magnitude of this term is given by summing over the Rooseland mean opacities of the inverse Primakoff process and direct decay to photons. For the former this is given by \cite{Cadamuro:2011fd}
\begin{equation}
    \kappa_a^{P}=\frac{\int_{m_a}^{\infty}d\omega\ \omega^3\beta\frac{\partial}{\partial T} \frac{1}{\exp(\omega/T)-1}}{\rho \int_{m_a}^{\infty}d\omega\ \omega^3\lambda_a\beta\frac{\partial}{\partial T} \frac{1}{\exp(\omega/T)-1}}
\end{equation}
where \(\omega\) now refers to the ALP energy and \(\beta\) is the ALP velocity. The ALP mean-free path \(\lambda_a\) is given by \cite{Cadamuro:2011fd}
\begin{align}
    \lambda_{a}^{-1}=\sum_{Z}n_Z\sigma_{Z}^{bc}(\omega).
\end{align}
Here \(\sigma_{Z}^{bc}(\omega)\) is the cross-section for the inverse Primakoff process, or \textit{back-conversion} (bc), for a target of charge \(Ze\) and \(n_Z\) refers to its number density. This cross-section is given by \cite{Dolan:2017osp}
\begin{equation}
    \sigma_{Z}^{bc}=\frac{2}{\beta^2}\sigma_{Z}^P(\omega) \, ,
\end{equation}
where $\sigma_{Z}^P$ is the production cross-section.
The contribution towards ALP Rooseland mean opacity due to decay can be calculated in a similar fashion, although only the high-mass limit ($m_a/T\gg1)$ is relevant to us, given by \cite{Raffelt_Energy_Transfer}
\begin{equation}
    \label{eq: ALP opacity}
    \kappa_{a}^D=\frac{(2\pi)^{7/2}}{45\rho} \bigg(\frac{T}{m_a}\bigg)^{5/2}\exp(m_a/T)\ \Gamma_{a}.
\end{equation}
The total ALP opacity is then \(\kappa_a=\kappa_a^P+\kappa_a^D\) \cite{Dolan:2017osp}. For MeV scale ALPs direct decay is the dominant contribution towards their energy transfer. The total radiative opacity $\kappa_{\mathrm{Rad}}$ is then given as
\begin{equation}
    \kappa_{\mathrm{Rad}}^{-1}=\kappa_{\gamma}^{-1}+\kappa_{a}^{-1},
\end{equation}
where $\kappa_{\gamma}$ is the photon opacity. Note that the ALP contribution to energy transport would only impact the structural evolution of a star if the affected stellar region is radiative rather than convective.

%% file: Section3/Sec3.tex
\section{Asymptotic Giant Branch Stars}
\label{sec: Sec 3}
\begin{figure}[t]
    \centering
    \includegraphics{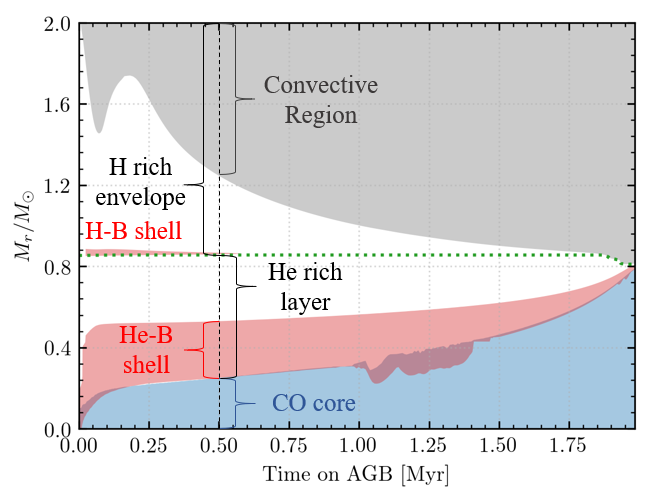}
    \caption{Kippenhahn diagram showing the evolution of the He-B and H-B shells (red), CO core (blue), convective zone (grey) of a $4M_{\odot}$ star throughout the E-AGB. The extent of these regions at time $0.5$~Myr are highlighted by the curly brackets, defined with respect to the radial mass coordinate $M_r$, i.e. the mass enclosed in a spherical shell of radius $r$. Also shown is the H/He discontinuity (green dotted line).}
    \label{fig: 4M Kippenhahn}
\end{figure}

Asymptotic giants are a class of cool, luminous star which have evolved beyond the phase of central helium burning. The asymptotic giant branch (AGB) is an evolutionary stage experienced only by stars in the approximate mass range of $0.8-8M_{\odot}$ \cite{2012sse..book.....K}, which are massive enough for helium burning to occur, but insufficiently massive to support non-degenerate carbon fusion. A comprehensive review can be found at \cite{AGBStarsBook}.

An AGB star has at its centre a core composed of carbon and oxygen, the products of helium fusion. Surrounding the core is a helium-rich layer at the base of which is a shell supporting helium-burning. This shell is a remnant of the previous phase of central He-B. The outer stellar envelope is composed primarily of hydrogen and hosts a convective layer which penetrates from the surface deep within the star and efficiently mixes its contents. A hydrogen-burning shell exists at the bottom of this, which has persisted from the end of the main sequence throughout central He-B.

The evolution of the inner $2M_{\odot}$ of a $4M_{\odot}$ star from the point of exhaustion of central helium is shown in the Kippenhahn diagram Figure \ref{fig: 4M Kippenhahn}. A Kippenhahn diagram depicts changes in stellar structure across evolutionary periods. At a given moment in time, the extent of stellar regions (e.g. the core, convective and burning regions) can be read along the vertical axis and are defined in terms of the radial mass coordinate $M_r$\footnote{Stellar structure equations are typically defined in terms of the radial mass coordinate, i.e. the mass interior to a spherical shell of radius $r$, rather than the radius itself. Spherical symmetry has been assumed.}. For example, when the star has been on the AGB for 0.5 Myr, the innermost $0.25M_{\odot}$, shown in blue, is occupied by the CO core. This is surrounded by the helium-rich zone which extends from $0.25M_{\odot}$ to $\approx0.85M_{\odot}$, with the He-B shell (red) occupying the bottom $0.25M_{\odot}$ of this. The entire region external to this is occupied by the outer hydrogen envelope. The convective layer extends as far down as $M_r\approx1.25M_{\odot}$, and a thin hydrogen-burning shell remains at $M_r\approx0.85M_{\odot}$. The dotted green line indicates the mass coordinate of the boundary between hydrogen- and helium-rich zones.

\begin{figure}[t]
    \centering
    \includegraphics{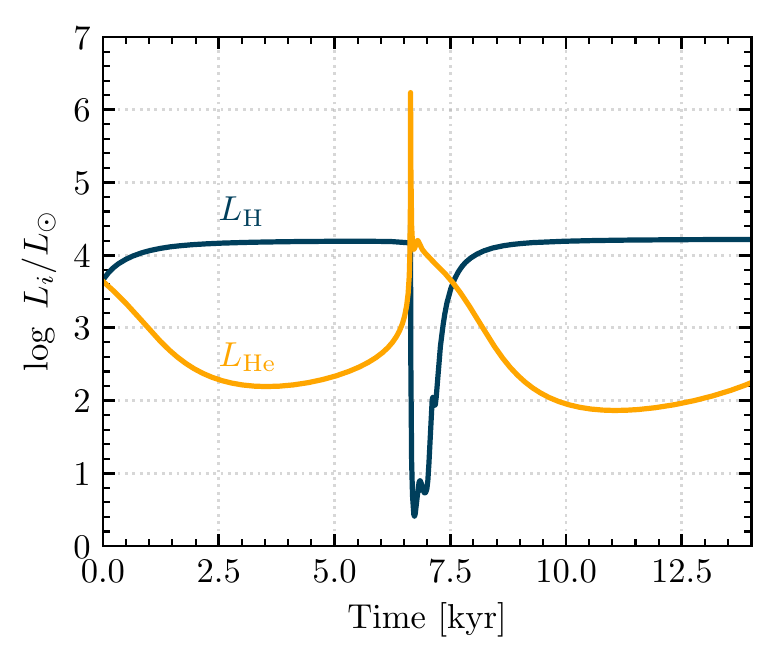}
    \caption{The evolution of the luminosities $L_{\mathrm{H}}$ (blue) and $L_{\mathrm{He}}$ (yellow) throughout a typical pulse cycle for a $4M_{\odot}$ star. Time has been set to zero when quiescent hydrogen burning begins.}
    \label{fig: 4M TP}
\end{figure}

\subsubsection*{The early-AGB:}
The onset of the AGB coincides with considerable structural change within the star. Once the He-B shell has been established surrounding the CO core, its substantial energy output prompts the expansion and cooling of the entire He-rich layer. Consequently, nuclear activity within the superior H-B shell is suppressed and, if $M_{\mathrm{init}}\gtrsim4M_{\odot}$, extinguished. What follows is a period of stable helium shell burning and CO core growth, known as the early-AGB (E-AGB).

For helium fusion to be sustained throughout the E-AGB, the He-B shell must gradually progress outward through the He-rich region. Throughout this process the He-B shell begins to thin, causing its temperature to increase and nuclear activity to intensify. Much of the associated energy-flux drives further expansion and cooling of the outer layers, enabling the convective zone to penetrate more deeply into the stellar envelope.

A critical point is reached when the convective zone reaches the H/He discontinuity ($\sim1.87$ Myr in Figure \ref{fig: 4M Kippenhahn}). In stars with $M_{\mathrm{init}}\lesssim4M_{\odot}$, the still functional H-B shell prevents any deeper incursion of the convective zone and the H/He discontinuity is left unaltered. However in more massive stars, where the H-B shell is dormant, the convective zone breaches the H/He discontinuity and delves into the helium-rich region below, dispersing its contents (namely helium and nitrogen) throughout the outer-envelope. This event is termed the second dredge-up\footnote{The first dredge-up occurs at the end of the main-sequence when a star approaches the red-giant branch.}. Notably the second dredge-up disperses a substantial amount of fuel for the He-B shell, which restricts CO core growth throughout the E-AGB.

\subsubsection*{The thermal pulsating-AGB:}
The E-AGB is brought to a close when the He-B layer approaches the H/He discontinuity and its supply of nuclear fuel dwindles. The associated decline in helium burning activity allows the outer-envelope to contract, reigniting the dormant hydrogen shell. Interestingly, the geometrically thin He-B is thermally unstable, which facilitates the development of pulsations within the star's outer layers. These thermal pulsations characterise the second phase of the AGB, the thermal pulsating-AGB (TP-AGB).

\begin{figure}[t]
\centering
\includegraphics{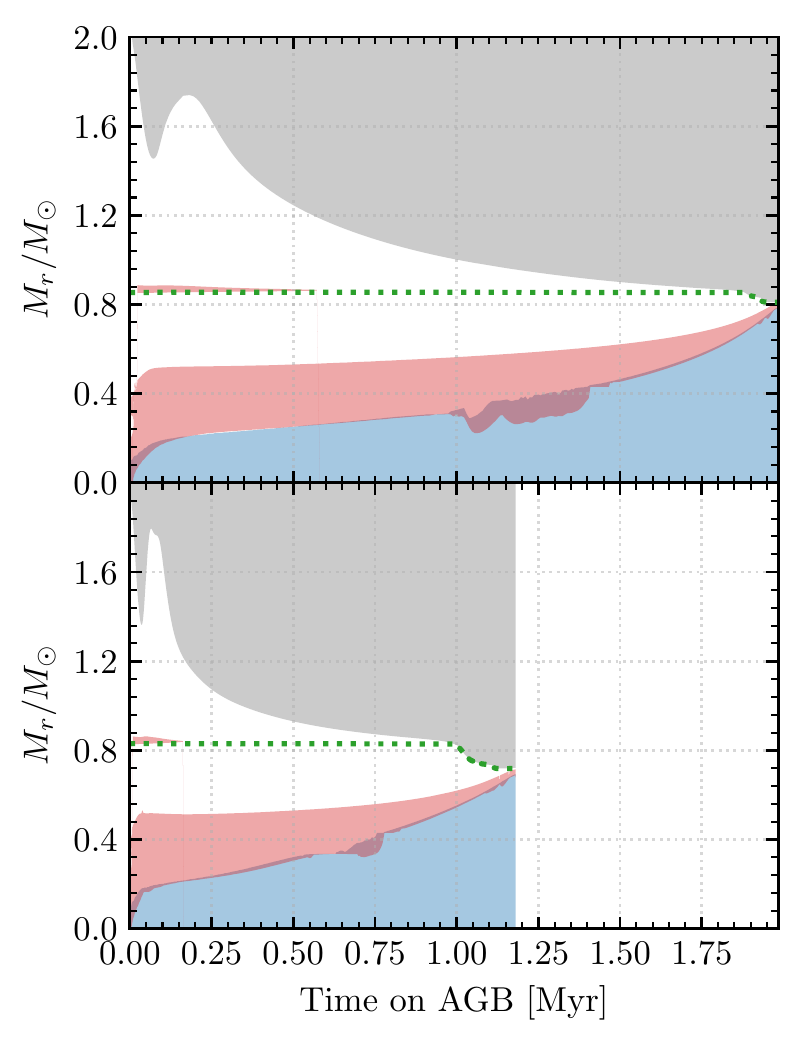}
\caption{Comparison between the E-AGB evolution of stellar structure for $4M_{\odot}$ stars with (bottom) and without (top) the inclusion of 10~keV ALPs with $g_{a\gamma\gamma}=0.63\times10^{-10}$~GeV$^{-1}$. All elements of the plot are defined in Figure \ref{fig: 4M Kippenhahn}.}
\label{fig: Low Mass AGB Comp}
\end{figure}
\begin{figure}[t]
    \centering
    \includegraphics{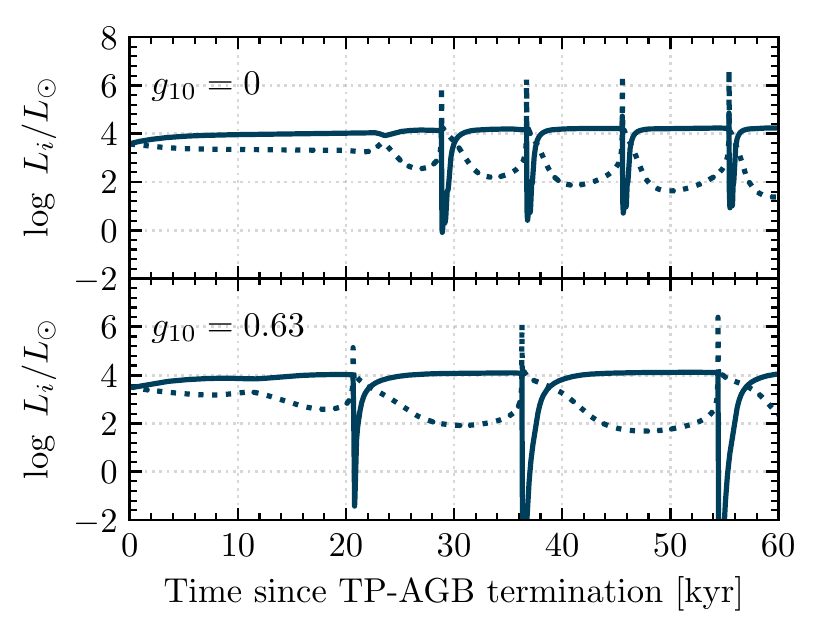}
    \caption{Comparison between the thermal pulses of the TP-AGB for $4M_{\odot}$ stars with $g_{10}=g_{a\gamma\gamma}/(10^{-10}\ GeV^{-1})=0.0$ (top) and 0.63 (bottom). The hydrogen and helium luminosities are indicated by solid and dotted lines respectively.}
    \label{fig: TP Comparison}
\end{figure}

A typical pulse cycle, illustrated in terms of the hydrogen (blue) and helium (yellow) luminosities, is shown in Figure \ref{fig: 4M TP} for the same $4M_{\odot}$ star whose evolution is illustrated in Figure \ref{fig: 4M Kippenhahn}. Sparse fuel supply in the He-B shell causes nuclear activity therein to dwindle, giving way to a long period of \textit{quiescent} hydrogen shell burning (the inter-pulse period). The helium produced during this time settles onto the helium-rich region below, increasing its mass and causing the pressure and temperature at its base to rise. 

Once the mass of this inter-shell region reaches a certain threshold, helium is re-ignited in an unstable event known as the \textit{helium shell flash}. Such flashes are brief, occurring on scales of $\mathcal{O}(1\ \mathrm{yr})$, and suppress hydrogen-shell burning. They are followed by a period of stable He-B which is sustained for a few hundred years. Once its fuel has been exhausted, nuclear activity within the He-B shell again diminishes, giving way to quiescent H-burning. The duration of the inter-pulse period varies with the core mass, with more massive cores supporting more rapid pulsations \cite{Paczynski-IP-MC-reln, Boothroyd-core-IP-reln}.

Many thermal pulsations occur during the TP-AGB, each of which produces a non-trivial amount of helium, carbon and oxygen which increase the masses of the CO and hydrogen-depleted cores (with boundary defined by the H/He discontinuity) outwards. These can be accompanied by further dredge-up events, in which the convective layer again penetrates into a region containing the ashes of helium-burning (the third dredge-up), which reduces the growth of the cores \cite{2004ApJ...605..425H}. The TP-AGB, and indeed the AGB itself, is ultimately halted by strong stellar wind, which progressively strips the outer envelope and leaves only the remnant white dwarf. In low-mass stars ($\lesssim3M_{\odot}$), the TP-AGB is sufficiently long for thermal pulses to contribute to the final stellar mass $M_{\mathrm{f}}$ by as much as 30\% \cite{Cummings_2019}. More massive stars, however, shed their envelopes much more rapidly and experience only marginal core growth during the TP-AGB \cite{10.1093/mnras/stt1034}. For the $4M_{\odot}$ star shown in Figure \ref{fig: 4M Kippenhahn}, the hydrogen-depleted core grows only from $0.81M_{\odot}$ to $0.825M_{\odot}$, an increase of approximately 2\%.

\subsection{Axion-like particles and the AGB}

To probe the impact of ALPs on the AGB, evolutionary simulations of $4M_{\odot}$ stars were computed for two choices of $m_a$ and $g_{a\gamma\gamma}$ values. We distinguish between the cases of low and high mass ALPs below. The first case corroborates the results presented in \cite{Dominguez}, though we include only the ALP-photon interaction. We then show that this behaviour persists to heavier ALPs.

\subsubsection*{Light ALPS:}

The E-AGB phase from the simulation with $g_{a\gamma\gamma}=0.63\times10^{-10}$ GeV$^{-1}$ and $m_a=10\ \mathrm{keV}$ is shown in the lower panel of Figure \ref{fig: Low Mass AGB Comp}. The upper panel contains the same phase given standard astrophysics alone. The inclusion of these ALPs within the simulation expedites the E-AGB.

The primary culprit for this is ALP-production and escape in the He-B shell. Energy loss within this region prompts it to contract and heat, which accelerates nuclear fuel consumption and causes the entire evolutionary phase to occur more rapidly. More intense nuclear burning also sparks a premature and deeper second dredge-up event, which displaces a greater mass of helium-rich material throughout the convective region. This naturally increases the surface abundances of the remnants of nuclear burning (He and $^{14}$N). A deeper dredge-up event also results in a smaller H-depleted core mass $M_c$, which reduces possible CO core growth during the E-AGB. This is precisely what is seen in our models, where the terminal E-AGB value of $M_c$ and $M_{CO}$ decreases from $0.81M_{\odot}$ and $0.79M_{\odot}$ to $0.72M_{\odot}$ and $0.69M_{\odot}$ respectively.

\begin{figure}[t]
    \centering
    \includegraphics[width=8.5cm]{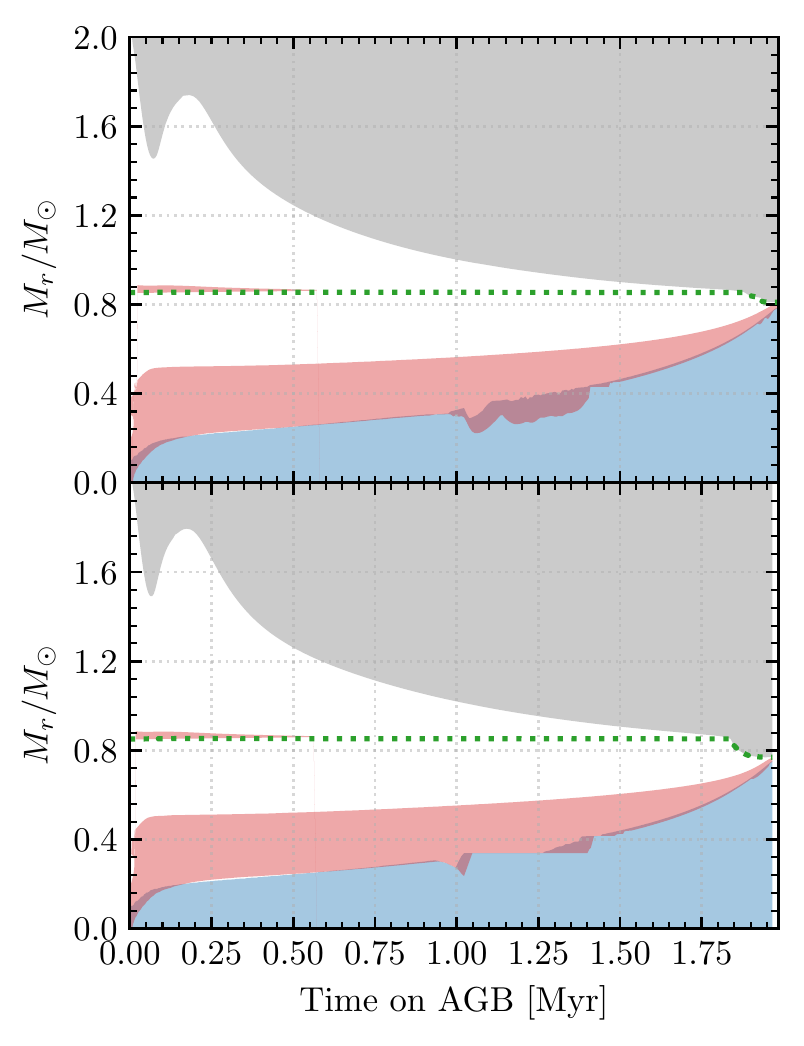}
    \caption{Comparison between the E-AGB evolution of stellar structure for $4M_{\odot}$ stars with (bottom) and without (top) the inclusion of 316keV ALPs with $g_{a\gamma\gamma}=10^{-9}$ GeV$^{-1}$. This choice of ALP parameters is well outside the HB star bound in Figure \ref{fig: ALP_param_space}.}
    \label{fig: 4M comparison}
\end{figure}

The addition of ALPs also disrupts the evolution of the TP-AGB. As shown in Figure \ref{fig: TP Comparison}, when ALPs are included in the model, the length of the inter-pulse period increases substantially. This is primarily caused by the decreased core mass at the onset of pulsation \cite{Dominguez}. Furthermore, energy-loss within the He-B shell during these pulses produces more extreme third dredge-up events. This produces a potentially observable signature, as the surface abundances of the products of helium-burning increase \cite{Dominguez}. Potential ramifications of lengthening the inter-pulse period as well as deeper third dredge-up events are discussed in Section \ref{sec: Section 5}.

\subsubsection*{Heavy ALPs:}

In order to investigate the impact of heavier ALPs we recompute the evolution of the $4M_{\odot}$ star with $m_a=316$ keV and $g_{a\gamma\gamma}=10^{-9}$ GeV$^{-1}$, well outside the region constrained by HB stars. The E-AGB evolution of this star is shown in Figure \ref{fig: 4M comparison}.

The production of heavier ALPs in the He-B shell is Boltzmann suppressed and consequently there is minimal reduction in the duration of the early asymptotic giant branch phase. Like the light ALP case, however, the second dredge-up penetrates more deeply into the helium-rich layer and, consequently, an observed reduction in $M_{c}$ and $M_{\mathrm{CO}}$ is retained ($0.81M_{\odot}$ and $0.79M_{\odot}$ to $0.77M_{\odot}$ $0.75M_{\odot}$), though this effect is of moderate strength only.

\begin{figure}[t]
\centering
\begin{minipage}[t]{.5\textwidth}
    \centering
    \captionsetup{width=0.9\textwidth}
    \includegraphics[width = 7.5cm]{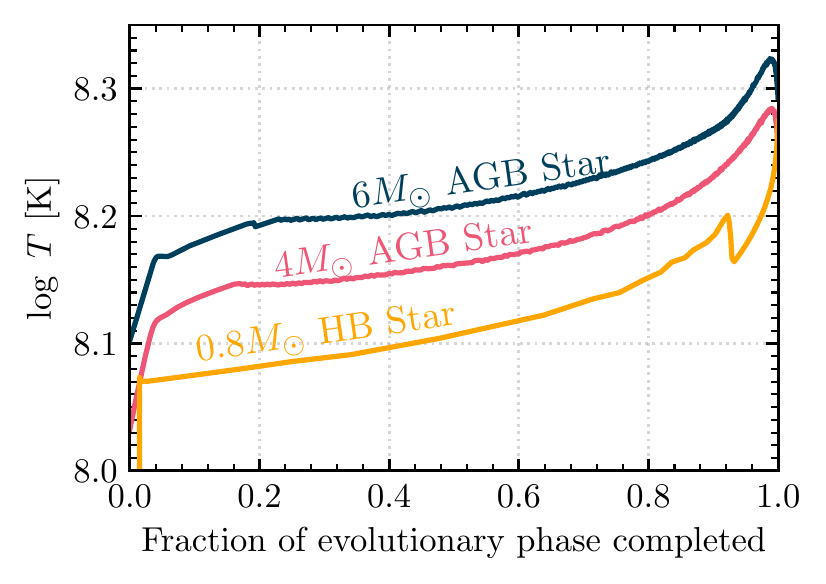}
    \caption{The evolution of temperature at the location of maximum helium burning in models of $4M_{\odot}$ and $6M_{\odot}$ AGB stars as well as a $0.8M_{\odot}$ horizontal branch star in the absence of ALPs. Evolution is depicted as a fraction of the total duration of the phase.}
    \label{fig: Temperature Evolution}
\end{minipage}%
\begin{minipage}[t]{.5\textwidth}
    \centering
    \captionsetup{width=0.9\textwidth}
    \includegraphics[width=7.5cm]{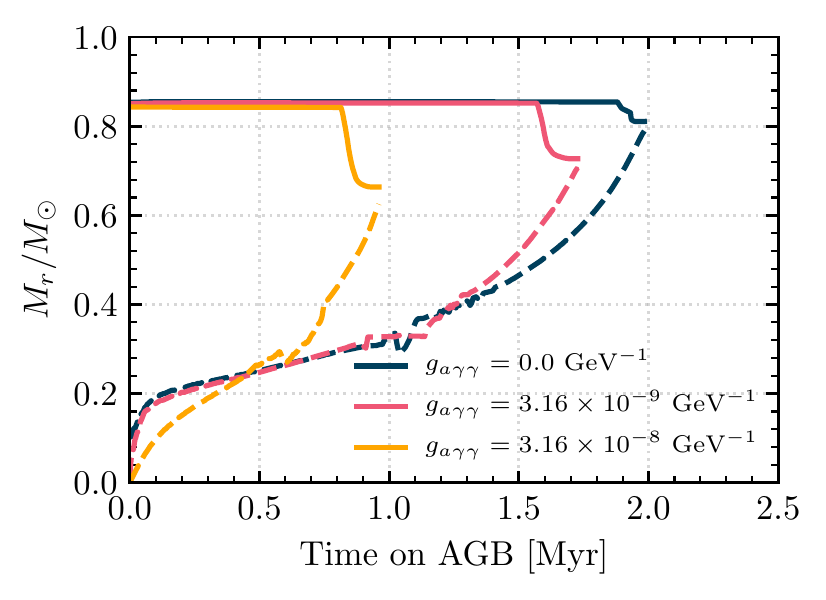}
    \caption{The mass coordinate of the H/He discontinuity (solid) and $M_{\mathrm{CO}}$ (dashed) in $4M_{\odot}$ stars for 316 keV ALPs with $g_{a\gamma\gamma}=3.16\times10^{-9}$ GeV$^{-1}$ (pink), $g_{a\gamma\gamma}=3.16\times10^{-8}$ GeV$^{-1}$ (yellow) and with the ALP-photon interaction switched off (dark blue).}
    \label{fig: Other tracks}
\end{minipage}
\end{figure}

The persistence of ALP effects on the second dredge-up can be understood by examining the evolution of temperature in the He-B layer during the E-AGB, indicated by the pink line in Figure \ref{fig: Temperature Evolution}. For the majority of the E-AGB the temperature of the shell is below $T=10^{8.2}\ \mathrm{K}\approx 1.6\times10^{8}\ \mathrm{K}$ and little energy-loss to ALP-production occurs (cf. Figure \ref{fig: Eps_masscomp}). As the layer thins, its temperature steadily increases, alleviating the Boltzmann suppression of ALP production. This triggers the positive-feedback loop, which facilitates a more intense spike in helium luminosity and the rapid establishment of a deep dredge-up event.

Figure \ref{fig: Other tracks} depicts the evolution of the H/He discontinuity (solid line) and $M_{\mathrm{CO}}$ (dashed) for 316 keV ALPs with couplings strengths of $g_{a\gamma\gamma}=3.16\times10^{-9}$ GeV$^{-1}$ (purple), $g_{a\gamma\gamma}=3.16\times10^{-8}$ GeV$^{-1}$ (yellow) and with ALP-production switched off (dark blue). As expected, increasing the value of  $g_{a\gamma\gamma}$ causes $M_{\mathrm{CO}}$ and $M_c$ to decrease. Notably, approximately the same reduction in CO core mass is obtained when both 10 keV ALPs with $g_{a\gamma\gamma}=0.63\times10^{-10}$ GeV$^{-1}$ and 316 keV ALPs with $g_{a\gamma\gamma}=3.16\times10^{-9}$ GeV$^{-1}$ are included in the model. However, the latter only shortens the E-AGB duration by 13\%, rather than 41\% in the case of the former. This confirms that heavy ALPs, which only become relevant towards the end of the E-AGB, nevertheless impact stellar structure significantly.

The He-B shells of more massive stars, which comprise the upper-end of the IFMR (e.g. the $6M_{\odot}$ star in Figure \ref{fig: Temperature Evolution}), reach higher temperatures still during the E-AGB. Such objects should therefore show even greater sensitivity to heavy ALPs. This is precisely what we see in our simulations, which predict a 16\% reduction $M_{\mathrm{CO}}$ and a 46\% reduction in E-AGB duration for a $6M_{\odot}$ star experiencing energy-loss to 316 keV ALPs with $g_{a\gamma\gamma}=10^{-9}$ GeV$^{-1}$, compared with 5\% and 1\% respectively for the $4M_{\odot}$ stars in Figure \ref{fig: 4M comparison}.

Compared with the He-B shell of AGB stars, the cores of the $0.8M_{\odot}$ HB stars simulated in \cite{Carenza:2020zil} are cooler across all stages of their respective evolutionary phases (see the yellow line in Figure \ref{fig: Temperature Evolution}). It is precisely this which gives AGB stars great potential to further constrain the cosmological triangle. Although the HB star bound has been consistently refined over time, any constraint we derive based on the established effects on $M_{\mathrm{f}}$ will simply be sensitive to heavier ALPs.

It should be noted that ALPs retain the capacity to elongate the inter-pulse period during the TP-AGB, as this is principally a function of $M_{c}$. However, this does not significantly influence the constraint derived in Section \ref{sec: Section 4} and as such discussion of its potential constraining power is deferred to Section \ref{sec: Section 5}. 

%% file: Section4/Sec4.tex
\section{The White Dwarf Initial-Final Mass Relation}
\label{sec: Section 4}

The IFMR relates the initial mass with which a star forms to the mass of the white dwarf into which it ultimately evolves. IFMRs calibrated to observation therefore provide a constraint on total mass loss throughout a stellar lifetime, as well as free parameters of stellar modelling \cite{KaliraiMAssLoss, Cummings_2019}. The IFMR is instrumental in age and distance determination in globular clusters, our understanding of supernovae rates \cite{Greggio-TypeIa}, galactic chemical evolution and the field white dwarf population \cite{Kalirai:2007tq}.

\subsection{Constraints on the IFMR}

\begin{figure}[t]
    \centering
    \captionsetup{width=0.9\textwidth}
    \includegraphics{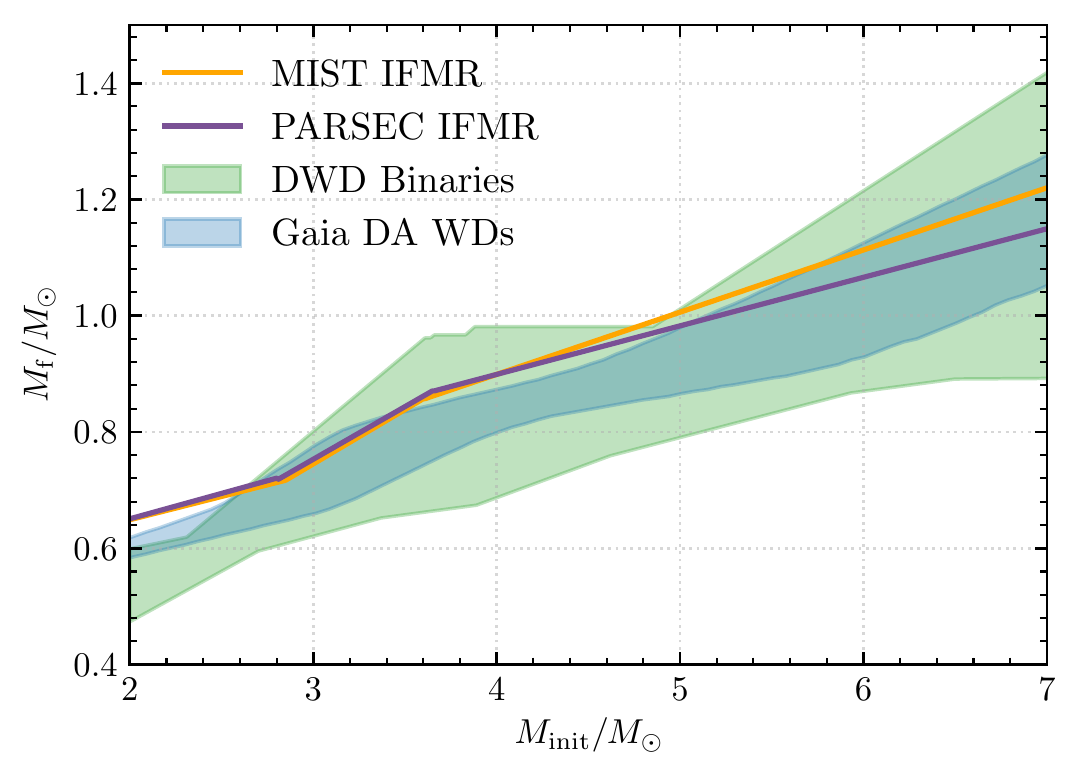}
    \caption{A sample of existing constraints on the IFMR, with the full range of the double white dwarf binary constraint with flexible breakpoints \cite{Andrews} shown in light green. Also included are the 95\% confidence interval of the constraint from 1100 Gaia hydrogen-rich white dwarfs within 100pc \cite{El-Badry} in blue and the star cluster semi-empirical IFMRs derived using stellar models and isochrones from MIST (yellow) and PARSEC (purple) \cite{Cummings_2018}}
    \label{fig: IFMR Example}
\end{figure}

Numerous constraints on the IFMR exist, the majority of which are derived using WDs in star clusters (e.g. \cite{Catalan2008, Kalirai_2008, Salaris_2009, Cummings_2018}). We provide a general description of the construction of these star cluster IFMRs, though a more detailed account can be found in \cite{Cummings_2018}.
\begin{enumerate}
    \item Spectroscopic analysis of the WDs enables their effective temperatures $T_{\mathrm{eff}}$ and surface gravity $\log(g)$ to be derived. These can be converted to white dwarf masses $M_{\mathrm{f}}$ and cooling age $\tau_C$ via application of theoretical white dwarf cooling models.
    \item If the age of the cluster $\tau_{\mathrm{SC}}$ which hosts the WD  is known, the progenitor lifetime can be determined as $\tau_{P}=\tau_{\mathrm{SC}}-\tau_{C}$. Cluster age determination is typically achieved through use of isochrone\footnote{ Isochrones are a complementary tool to the evolutionary models discussed thus far. While stellar tracks report information pertaining to the evolution of a single star with given initial mass \(M_{\textrm{init}}\) and metallicity $Z$, isochrones detail the properties of a cluster of items at a fixed age, as a function of their mass. Implicit in their use is the assumption that all objects described by a single isochrone have formed out of the same homogeneous gas cloud, and consequently have an identical composition.} fitting.
    \item Finally stellar evolution models of appropriate metallicity are used to determine the initial stellar mass $M_{\mathrm{init}}$ associated with the progenitor lifetime $\tau_P$. Repeating this process over an entire sample of WDs yields an IFMR calibrated to observation.
\end{enumerate}
The central values of two star-cluster IFMRs derived in \cite{Cummings_2018} are shown in Figure \ref{fig: IFMR Example}. In this analysis two different sets of isochrones and stellar evolution models were used, which produces the observed differences at large values of $M_{\mathrm{init}}$. These are the PARSEC isochrones \cite{ParsecIsochrones}, computed from the Padova stellar evolution models, as well as those of \texttt{MESA} Isochrones and Stellar Tracks (MIST) \cite{MIST0, MIST1} which are based on \texttt{MESA} simulations.

A large source of uncertainty in these semi-empirical IFMRs arises from the determination of cluster ages, which can vary significantly between stellar models with different treatments of rotation and core overshoot (see Appendix \ref{sec: AppC}) \cite{Andrews}. As our principal aim is to constrain physics beyond the Standard Model, the use of an IFMR constraint which removes much astrophysical uncertainty is desirable.

Recently there has been interest in finding complementary constraints on the IFMR. In \cite{El-Badry} an empirical measurement of the IFMR was sought via analysis of a sample of 1100 WDs from the \textit{Gaia} Data Release 2 \cite{Gaia1, Gaia2}. This is shown in blue in Figure \ref{fig: IFMR Example}. However it was argued in \cite{Cummings_2018} that, by restricting their data to white dwarfs which have previously been spectroscopically identified, non-trivial selection biases are introduced. Furthermore, their derived IFMR is sensitive to the choice of initial-mass function (IMF) for large initial masses, precisely where ALP effects are most significant.

An appropriate IFMR constraint for our purposes was derived in \cite{Andrews} from a sample of 14 wide double white dwarf binary systems of solar metallicity. Like their counterparts in open clusters, the masses and cooling times of these WDs can be determined by application of theoretical WD cooling models. Instead of relying on the computation of absolute progenitor lifetime, however, only the relative lifetime $\Delta\tau_P=\tau_{P}^{(1)}-\tau_{P}^{(2)}$ is necessary. Consequently, the constraint derived from this analysis is independent of cluster ages. As these binaries are wide, we can assume that they have evolved independently of one another as single stars. This detail is crucial, as binary stellar evolution modelling is beyond the scope of this work.

The constraint \cite{Andrews} was determined in the following manner.
\begin{enumerate}
    \item Spectroscopic analysis of the WD atmospheres when combined with synthetic WD cooling tracks enabled the determination of their final masses $M_{\mathrm{f}}^{(1)}$, $M_{\mathrm{f}}^{(2)}$ and cooling times $\tau_{C}^{(1)}$, $\tau_{C}^{(2)}$.
    \item The relative cooling time $\Delta\tau_{C}=\tau_{C}^{(1)}-\tau_{C}^{(2)}$ can then be determined. As binary companions, each pair of stars can be assumed to be the same age. Consequently the difference in progenitor lifetime is given by $\Delta\tau_P'=-\Delta\tau_C$.
    \item An initial parametric model for the IFMR is then assumed, modelled as a three-piece linear relation, which allows estimates for initial masses $M_{\mathrm{init}}^{(1)}$, $M_{\mathrm{init}}^{(2)}$ to be determined for each binary WD.
    \item Stellar evolution models can then be used to convert each initial mass to a theoretical progenitor lifetime $\tau_{P}^{(1)}$, $\tau_{P}^{(2)}$. How well the difference between these $\Delta\tau_P$ matches the observed $\Delta\tau_P'$ is used to define the likelihood that the parametric model of the IFMR matches the observational data.
    \item When this is iterated upon, the best-fit parametric model can be determined.
\end{enumerate}
In \cite{Andrews} this was repeated many times to derive a posterior sample of semi-empirical IFMRs.

For the majority of \cite{Andrews} breakpoints at $2M_{\odot}$ and $4M_{\odot}$ are assumed in the three-piece fit. The motivation for this choice is physical. Stars with $M_{\mathrm{init}}\lesssim2M_{\odot}$ experience a degenerate helium-flash, while in the range $2M_{\odot}\lesssim M_{\mathrm{init}}\lesssim4M_{\odot}$ helium burning proceeds in a stable, non-degenerate convective core \cite{1976ApJS...32..367S}. If $M_{\mathrm{init}}\gtrsim4M_{\odot}$, a second dredge-up event can occur, which flattens the IFMR. The posterior sample for this three-piece fit  was made available by the authors of \cite{Andrews}.

When the values of the breakpoints are allowed to vary, however, a wider spread of IFMRs is obtained, particularly for high initial stellar masses. Given that the value of these breakpoints vary in the literature (e.g. in the MIST IFMR of \cite{Cummings_2018}, upper and lower breakpoints of $2.85M_{\odot}$ and $3.6M_{\odot}$ respectively produce the best fit), we conservatively favour this less restrictive constraint. The posterior sample in this case was not made available, which prevents a probabilistic interpretation of the results. Consequently, we show the entire range of IFMRs allowed in Figure \ref{fig: IFMR Example}.

A degree of tension exists between constraints for low initial masses. When $M_{\mathrm{init}}\lesssim2.6M_{\odot}$, the central fit for the MIST and PARSEC IFMRs is approximately $0.02M_{\odot}$ higher than the upper boundary of the Gaia constraint. This discrepancy is worse for DWD binaries, the upper limit of which falls approximately $0.04M_{\odot}$ below the central values of the cluster IFMRs when $M_{\mathrm{init}}\lesssim2.4M_{\odot}$.

Multiple sources of this tension have been suggested, including errors in star cluster ages or the presence of unresolved binaries in the WD samples \cite{Andrews}. This first possibility in particular motivates our choice of the DWD constraint, which is independent of star cluster ages and consequently more general. Clearly this constraint is far less restrictive than both the cluster IFMRs and the Gaia bound, particularly for the large initial masses which are sensitive to the influence of ALPs. As we shall see, however, even when a conservative approach is adopted, a substantial region of the ALP cosmological triangle can be ruled out.

\subsection{Comparing theory and observation}
\label{subsec: comparing theory and observation}
\begin{figure}[t]
    \centering
    \includegraphics[width = 15cm]{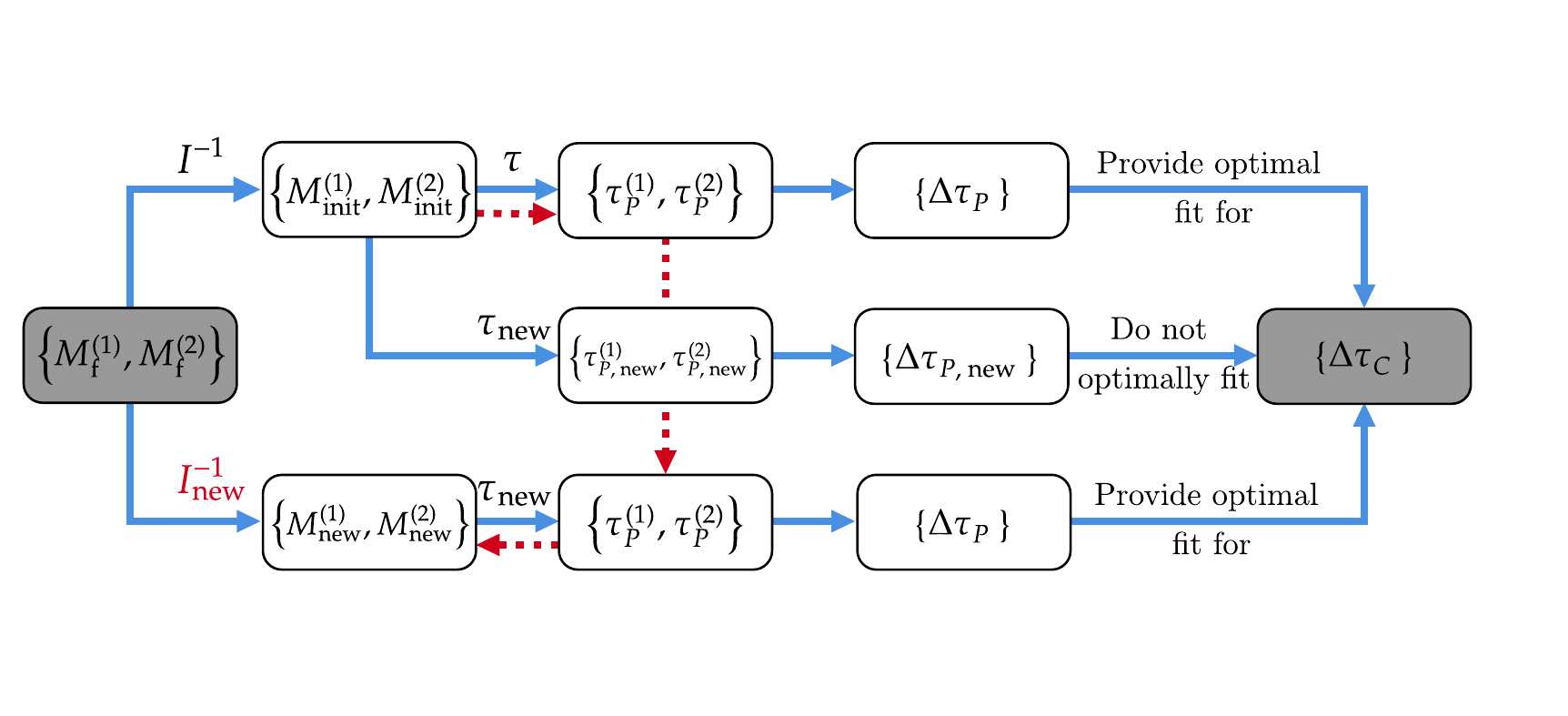}
    \caption{Diagram illustrating the role of progenitor lifetimes in the derivation of the wide DWD binary constraint of \cite{Andrews}. The top row indicates how the inverse IFMRs $I^{-1}$ supplied in the posterior distribution of \cite{Andrews} were optimised using progenitor lifetime function $\tau(M)$. The second row depicts how the use of modified progenitor lifetimes, given by $\tau_{\mathrm{new}}$, affect this analysis. The final row illustrates how these changes can be overcome by the identification of a updated inverse IFMR $I^{-1}_{\mathrm{new}}$. The dotted red lines indicate how $M_{\mathrm{new}}=I^{-1}_{\mathrm{new}}(M_{\mathrm{f}})$ may be related to $M_{\mathrm{init}}=I^{-1}(M_{\mathrm{f}})$.}
    \label{fig: Progenitor diagram}
\end{figure}
The constraint \cite{Andrews} has been selected to ease the comparison of our simulations with observation. However, some dependence on theoretical calculations is retained in \cite{Andrews} and we now discuss their impact.

\subsubsection*{Progenitor Lifetimes}
Given ALP-production drains energy from nuclear burning regions, thereby accelerating evolutionary timescales, it is important to discuss the role which stellar evolution models play in the semi-empirical constraint \cite{Andrews}. Note that a comprehensive discussion of the effects of these models would demand the re-evaluation of this constraint. This is beyond the scope of this work and consequently we present a simpler discussion below.

The role of these progenitor lifetimes is illustrated in the top row of the diagram in Figure \ref{fig: Progenitor diagram}. The constraint begins with a set of binary white dwarf masses $\{M_{\mathrm{f}}^{(1)},\ M_{\mathrm{f}}^{(2)}\}$ and associated difference in cooling times $\{\Delta\tau_C\}$ (shown in grey). An inverse IFMR $I^{-1}$ relates these white dwarf masses to a set of initial masses $\{M_{\mathrm{init}}^{(1)},\ M_{\mathrm{init}}^{(2)}\}$. These initial masses are then converted into their progenitor lifetimes $\{\tau_P^{(1)},\ \tau_P^{(2)}\}$ via the application of a function $\tau(M)$, derived by interpolating over the theoretical predictions of stellar simulations. The set of progenitor lifetime differences $\{\Delta\tau_{P}\}$ can then be calculated and compared with the observed set cooling time differences $\{\Delta\tau_C\}$. The inverse IFMR $I^{-1}$ (or equivalently IFMR) which gives the best fit between the magnitudes of these differences is said to be optimal.

Suppose, however, that the stellar simulations are systematically incorrect, due either to important input physics which has been neglected or the presence of ALPs in the model, and associate a new progenitor lifetime function $\tau_{\mathrm{new}}$ with these updated models. The effects of the utilisation of $\tau_{\mathrm{new}}$ are illustrated in the second line of Figure \ref{fig: Progenitor diagram}. Again the optimised inverse IFMR $I^{-1}$ is applied to the set of white dwarf masses, generating the same set of initial masses as in the first case. However, when new progenitor lifetimes are calculated using $\tau_{\mathrm{new}}$ they result in differences $\{\Delta\tau_{P,\ \mathrm{new}}\}$ which no longer provide the best fit for the set $\{\Delta\tau_C\}$. Consequently a new inverse IFMR $I_{\mathrm{new}}^{-1}$ must be found which maps the white dwarf masses to a set of initial masses $\{M_{\mathrm{new}}^{(1)},\ M_{\mathrm{new}}^{(2)}\}$, which return the original progenitor lifetimes  $\{\tau_P^{(1)},\ \tau_P^{(2)}\}$ (and consequently the optimal set of progenitor life differences $\{\Delta\tau_{P}\}$) upon application of $\tau_{\mathrm{new}}$. This scenario is shown in the final line of Figure \ref{fig: Progenitor diagram}.

We can estimate the impact of changing progenitor lifetimes by considering the relative size of the variable $M_{\mathrm{new}}=I_{\mathrm{new}}^{-1}(M_{\mathrm{f}})$ in terms of the initial mass given by the original optimised inverse IFMR $M_{\mathrm{init}}=I^{-1}(M_{\mathrm{f}})$. From the dashed lines in Figure \ref{fig: Progenitor diagram} it is evident that $M_{\mathrm{new}}$ is given by
\begin{equation}
    M_{\mathrm{new}}=\tau_{\mathrm{new}}^{-1}(\tau(M_{\mathrm{init}})).
    \label{eq: Mnew}
\end{equation}
If the neglected input physics increases progenitor lifetimes (i.e. $\tau_{\mathrm{new}}/\tau>1$ for all initial masses), $M_{\mathrm{new}}$ will be larger than $M_{\mathrm{init}}$\footnote{Progenitor lifetimes decrease with increasing initial mass.}, and the new inverse IFMR will be lower and flatter than the original. Conversely, if the progenitor lifetimes are shorter (e.g. due to the presence of energy loss to ALP production), the new initial masses will be smaller than the original and the updated posterior distribution of IFMRs will be higher and steeper than their counterparts in the set provided in \cite{Andrews}. The opposite will occur if $\tau_{\mathrm{new}}/\tau<1$ for all initial masses.

The constraint derived in the following sections primarily concerns ALPs with masses greater than 100 keV. The production of such ALPs is Boltzmann suppressed during central hydrogen- and helium-burning and, as such, their inclusion in stellar models shortens progenitor lifetimes only marginally ($\tau_{\mathrm{new}}/\tau$ decreases approximately linearly from 0.99 for $2M_{\odot}$ stars to 0.97 for $8M_{\odot}$). Using equation \ref{eq: Mnew} we find that the value of $M_{\mathrm{new}}/M_{\mathrm{init}}$ decreases approximately linearly from 0.994 (a 0.6\% reduction) for $2M_{\odot}$ to 0.985 for $8M_{\odot}$ (a 1.5\% reduction). The impact this reduction has on any two members of the posterior distribution of \cite{Andrews} will differ. Consequently we estimate the effect of neglecting ALP effects on stellar lifetime by applying this shift in $M_{\mathrm{init}}$ to all IFMRs with fixed breakpoints in the posterior distribution and identifying the average shift between these new IFMRs and the originals (the distribution of IFMRs with unfixed breakpoints was not made publicly available). When this approach is taken, we find that the posterior distribution is shifted upward by $0.002M_{\odot}$ and $0.007M_{\odot}$ in the $2$-$4M_{\odot}$ and $4$-$8M_{\odot}$ initial mass ranges respectively. 

These values clearly indicate that effects which reduce progenitor lifetimes, such as the inclusion of energy loss to ALPs, shift the IFMRs in the constraint of \cite{Andrews} upwards. For 100 keV ALPs and above, this magnitude of this effect is small and it will not significantly influence our results. ALPs with masses below a few keV, the production of which is not Boltzmann suppressed, can substantially reduce the magnitude of progenitor lifetimes. For example, 1 keV ALPs with a coupling strength of $g_{a\gamma\gamma}=1.58\times10^{-10}$ GeV$^{-1}$ reduce lifetimes by between 6-8\% for stars in the $4$-$8M_{\odot}$ mass range and as much as 12\% for $2M_{\odot}$ stars. If reductions of this scale were taken into account in the derivation of the constraint \cite{Andrews}, it would improve the constraining power of our bound. This, however, is beyond the scope of the present work.

It should be noted that, in the analysis presented in \cite{Andrews}, altering elements of input physics was found to yield comparable changes in progenitor lifetimes to those identified for 100 keV ALPs. These were deemed unimportant by the authors owing to the large uncertainty in stellar lifetimes generated by, for example, error in white dwarf mass measurement. The above discussion seems to support this conclusion.

\subsubsection*{White dwarf evolutionary models}
\begin{figure}[t]
    \centering
    \includegraphics{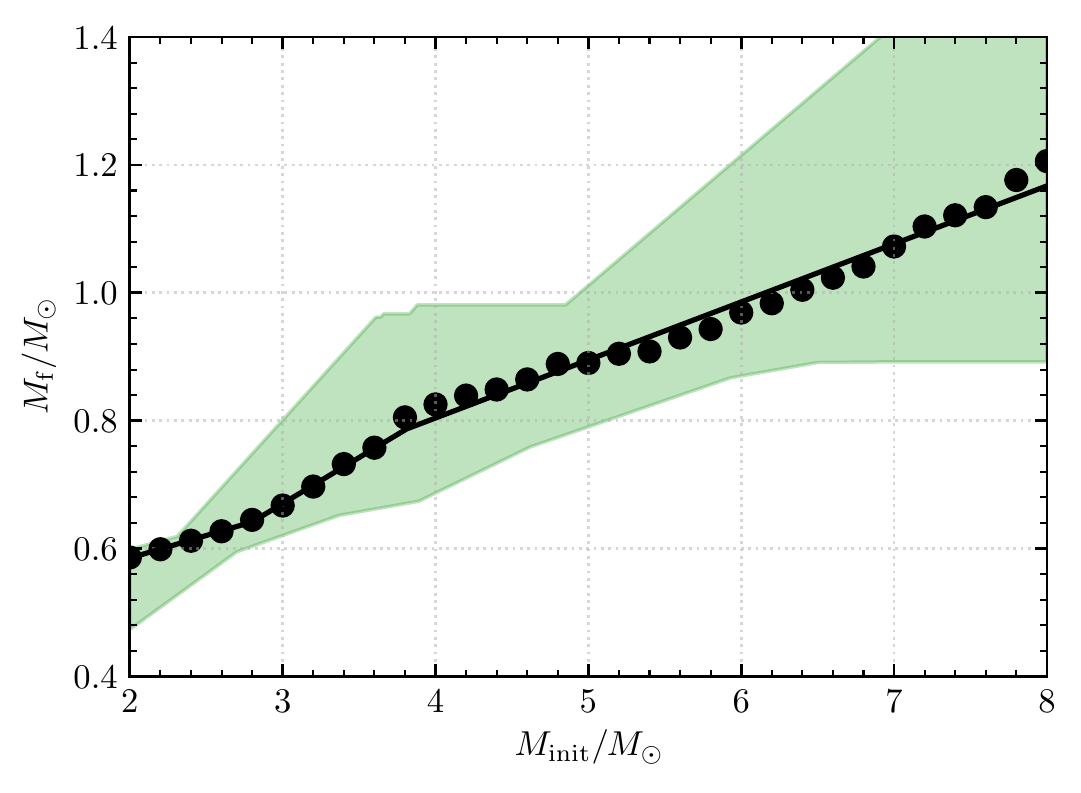}
    \caption{The $M_{\mathrm{f}}$ values from our simulations without ALPs for initial masses in $0.2M_{\odot}$ intervals between $2-8M_{\odot}$ (black points).  A three-piece linear fit with flexible breakpoints has been applied (solid black line). The double white dwarf binary constraint \cite{Andrews} is provided for comparison (green).}
    \label{fig: IFMR fit}
\end{figure}

The synthetic white dwarf evolutionary models of \cite{2001PASP..113..409F} are also used in the derivation of \cite{Andrews} to generate the necessary cooling times. White dwarf evolution is a fairly well-understood cooling process dominated in phases by neutrino emission, gravothermal settling and crystallisation (for a detailed explanation see \cite{2001PASP..113..409F, Tremblay2019}). As such, cooling models are often used to determine the ages of stellar populations (see e.g. \cite{Bedin}) and place limits on neutrinos \cite{Wignet2004ApJ, Hansen_2015}.

However,  anomalies in white dwarf cooling may exist. For example, tension exists between the observed and theoretical white dwarf luminosity functions (WDLFs) in certain stellar populations, which probe WD cooling  \cite{Isern1992, Isern:2008nt, Isern:2018uce, Bertolami:2014wua}. Interestingly, this tension can be explained by additional energy-loss to a DFSZ \cite{DINE1981199, Zhitnitsky:1980tq} type axion with $m_a\sim5$ meV. The WDLF, however, is sensitive to other astronomical properties including the initial mass function, star formation histories and the initial-final mass relation itself. Furthermore some WDLFs do not favour the existence of any additional cooling mechanism (e.g. that of \cite{Harris_2006} studied in \cite{Bertolami:2014wua}). Therefore until theoretical and observational uncertainties surrounding the WDLF improve, this remains only a hint towards the existence of such an axion.

Naturally, the existence of enhanced white dwarf cooling would require synthetic models to be updated. Though this could affect constraints such as \cite{Andrews}, it is worth noting that errors pertaining to cooling times are taken into account, with their most significant contribution coming from white dwarf mass measurement uncertainties.


With these factors considered, an initial set of stellar simulations was computed in $0.2M_{\odot}$ increments over the $2-8M_{\odot}$ range without energy-loss to ALPs. These were allowed to run from the pre-Main Sequence until the termination of the AGB, where mass loss has reduced the outer envelope to 1\% of the total stellar mass. The adopted input physics for these simulations mimics that used to calculate the MIST isochrones \cite{MIST0, MIST1}.

The initial and final stellar masses for these simulations are shown in as the black points in Figure \ref{fig: IFMR fit}, alongside a theoretical IFMR derived by applying a three-piece linear fit, where the breakpoints remained unfixed. While this fits well within the DWD binary bound, it is considerably lower than the Gaia, MIST and PARSEC IFMR constraints. This discrepancy is well documented (see e.g. \cite{Cummings_2019}) and can be somewhat mitigated by the inclusion of rotation in the stellar models (Section \ref{subsec: rotn}) and core overshoot (Appendix \ref{sec: AppC}).

\subsection{Axion-like particles and the IFMR}
\label{sec: ALPs and the IFMR}

\begin{figure}[t]
\centering
\begin{minipage}[t]{.5\textwidth}
    \centering
    \captionsetup{width=0.9\textwidth}
    \includegraphics[width=7.5cm]{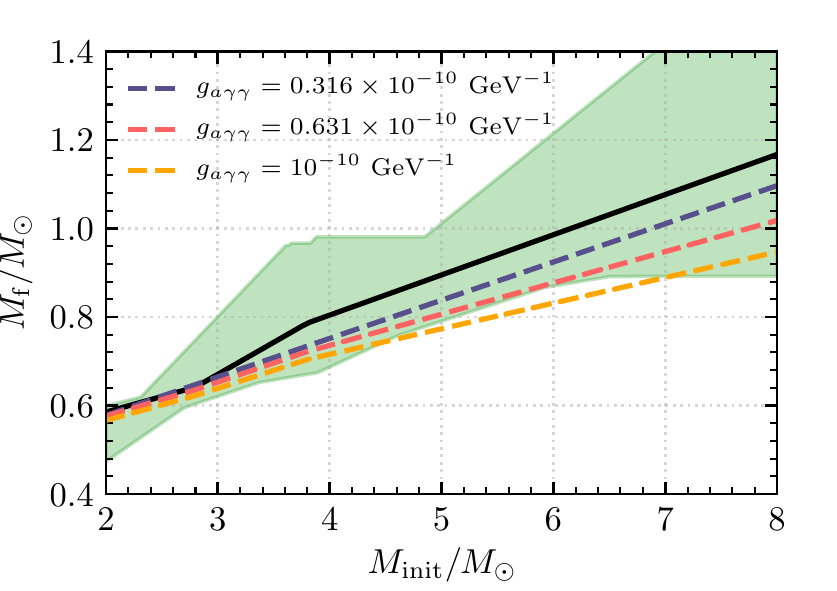}
    \caption{The three-piece IFMRs generated from our \texttt{MESA} simulations given the inclusions of energy-loss to 10 keV ALPs with $g_{a\gamma\gamma}=0.316\times10^{-10}$ GeV$^{-1}$ (purple), $0.631\times10^{-10}$ GeV$^{-1}$ (pink) and $10^{-10}$ GeV$^{-1}$ (yellow). The full range of the constraint \cite{Andrews} is included (green), along with our three-piece IFMR given standard astrophysics alone (black).}
    \label{fig: IFMR Run1}
\end{minipage}%
\begin{minipage}[t]{.5\textwidth}
    \centering
    \captionsetup{width=0.9\textwidth}
    \includegraphics[width=7.5cm]{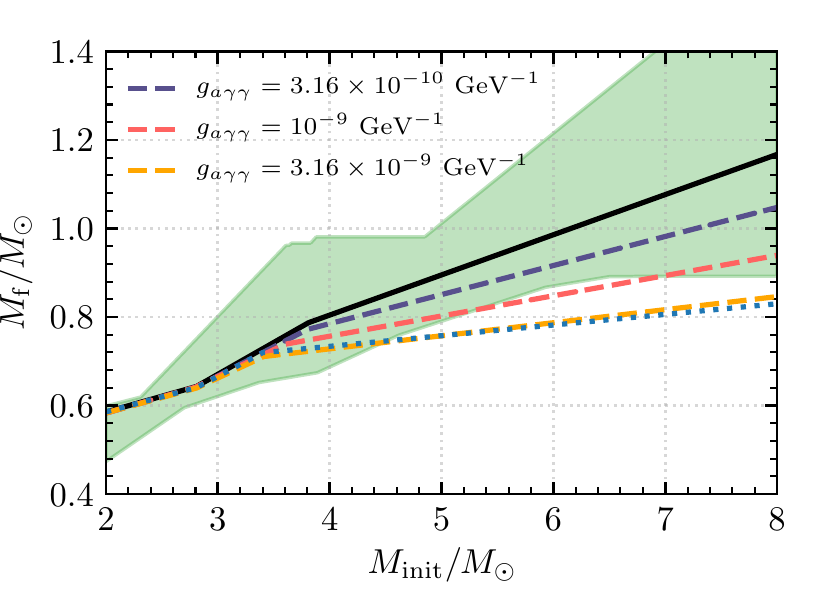}
    \caption{The three-piece IFMRs generated from our \texttt{MESA} simulations given the inclusions of energy-loss to 316 keV ALPs with $g_{a\gamma\gamma}=3.16\times10^{-10}$ GeV$^{-1}$ (purple), $10^{-9}$ GeV$^{-1}$ (pink) and $3.16\times10^{-9}$ GeV$^{-1}$ (yellow).  The full range of the constraint \cite{Andrews} is included (green), along with our three-piece IFMR given standard astrophysics alone (black). The dotted blue line indicates the IFMR given 630 keV ALPs with $g_{a\gamma\gamma}=10^{-5}$ GeV$^{-1}$. This choice of parameters is within the cosmological triangle.}
    \label{fig: IFMR Run2}
\end{minipage}
\end{figure}

To quantify the effects of axion-like particles on the IFMR, the series of simulations detailed in Section \ref{subsec: comparing theory and observation} were repeated including energy-loss to ALP-production. Initially 10~keV ALPs with $g_{a\gamma\gamma}=0.316\times10^{-10}$ GeV$^{-1}$ were considered and a new set of $M_{\mathrm{f}}$ values generated. The same three-piece fit specified in Section \ref{subsec: comparing theory and observation} was applied, the results of which are shown in Figure \ref{fig: IFMR Run1}. The departure from standard astrophysics due to deeper dredge-up events becomes evident for $M_{\mathrm{init}}\gtrsim3M_{\odot}$. This effect becomes more stark when larger values of $g_{a\gamma\gamma}$ are adopted. Ultimately, for $g_{a\gamma\gamma}=0.794\times10^{-10}$ GeV$^{-1}$, the theoretical IFMR falls outside the DWD constraint and this choice of ALP parameters can be excluded.

 Figure \ref{fig: IFMR Run2} depicts the results of repeating this analysis for 316~keV ALPs. The addition of ALP energy-loss again flattens the IFMR, however, this effect is considerably stronger in the $4$-$8M_{\odot}$ range, owing to the increased temperatures of their He-B shells (see Figure \ref{fig: Temperature Evolution}). Note that even the choice of parameters $g_{a\gamma\gamma}=3.16\times10^{-9}$~GeV$^{-1}$ associated with lowest IFMR in Figure \ref{fig: IFMR Run2} is unconstrained by the HB star bound. Also included is the three-piece IFMR generated if the ALP parameters $m_a=10^{5.8}\approx630$ keV and $g_{a\gamma\gamma}=10^{-5}$ GeV$^{-1}$ are chosen. Such ALPs lie within the cosmological triangle. Clearly, the IFMR is sensitive to ALPs from well within this region.

In order to construct a constraint, theoretical IFMRs were generated at logarithmically spaced intervals in $m_a$ and $g_{a\gamma\gamma}$. For each ALP mass, the smallest value of $g_{a\gamma\gamma}$ associated with an excluded theoretical IFMR was recorded. For some of these apparently excluded points the ALP will decay inside the nuclear burning region, so that energy loss does not result. While our simulations do not explicitly incorporate these effects, we take this into account below to derive a bound on the ALP parameter space.

\subsection{ALP-decay in AGB stars}
Any constraint derived from ALP influence on the He-B shell of AGB stars would not extend to arbitrarily high values of $g_{a\gamma\gamma}$, as ALP-decay becomes an important factor in that region. In order to quantify this, we follow the example of \cite{Carenza:2020zil}, who recently addressed ALP-decay in their HB star bound. In their treatment it was argued that the energy-loss constraint should relax when the decay-length
\begin{equation}
    \lambda_a=5.7\times10^{-5}g_{a\gamma\gamma}^{-2}m_{\mathrm{keV}}^{-3}\frac{\omega}{m_a}\sqrt{1-\bigg(\frac{m_a}{\omega}\bigg)^2} R_{\odot}
\end{equation}
falls below the HB star core radius $R_c\approx0.03R_{\odot}$, where $m_{\mathrm{keV}}=(m_a/1\ \mathrm{keV})$. Though ALPs will continue to contribute towards energy transfer after this (Section \ref{sec: Section 2}), this is sub-leading to convection which is dominant in the cores of such stars.

\begin{figure}[t]
    \centering
    \includegraphics{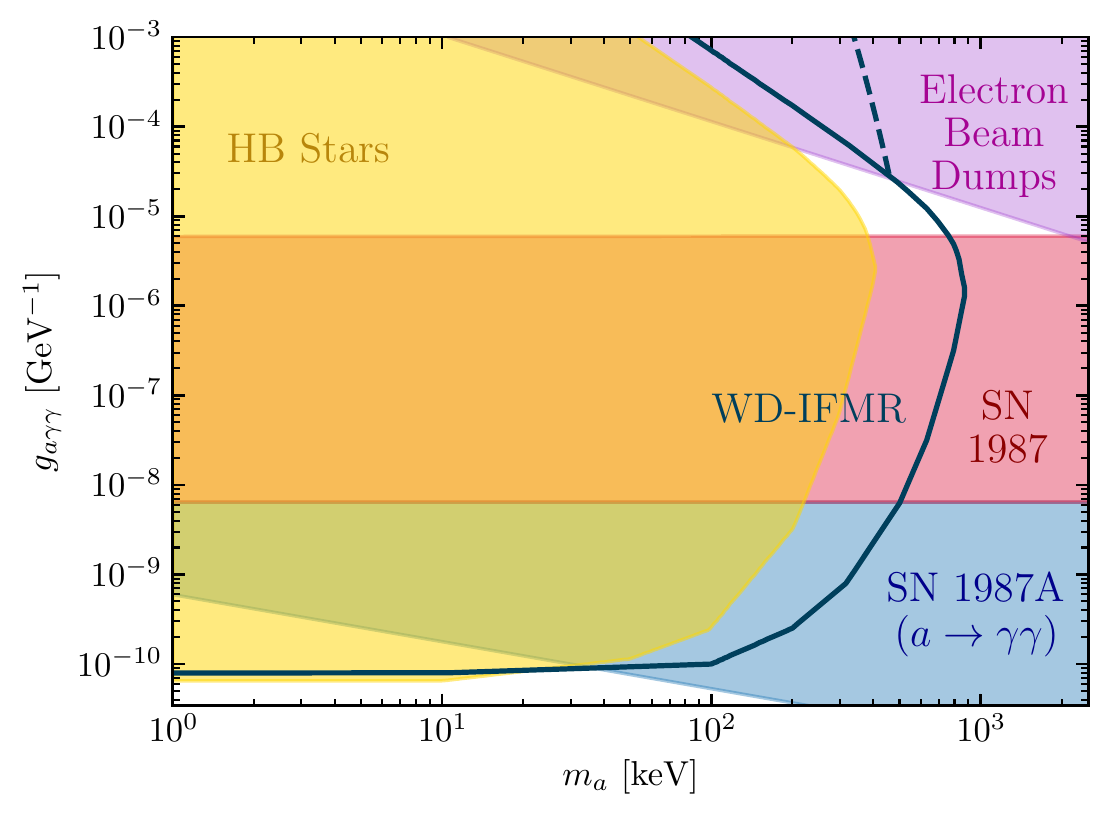}
    \caption{A suite of constraints in the keV-MeV region of the ALP plane. Our preliminary constraint including the impact of ALP-decay, is given by the solid dark blue line. The upper boundary of the nominal region in which ALPs are relevant for energy transfer in AGB stars is indicated by the dashed dark blue line. See text below for discussion.}
    \label{fig: ALP-IFMR Constraint}
\end{figure}

We similarly argue that the fundamental criterion for the energy loss argument, that energy be removed from a region undergoing nuclear burning, is no longer met in the He-B shell of AGB stars when $\lambda_a$ falls below the shell thickness $R_{\mathrm{He}}\approx0.007R_{\odot}$. Taking this into account yields the solid dark blue line in Fig.~\ref{fig: ALP-IFMR Constraint}.

Unlike HB cores, however, the He-B shell of AGB stars is radiative rather than convective. Consequently, ALPs can be expected to influence AGB stellar structure beyond the boundary of this constraint, though a thorough treatment of this requires dedicated simulations which include ALP contributions to energy transfer, which is beyond the scope of this work.

However, we are able to identify a region of parameter space in which ALPs still influence AGB structure. This is achieved by following the example of \cite{Raffelt_Energy_Transfer} and insisting that photons contribute dominantly towards radiative energy transfer in the He-B shell ($\kappa_a>\kappa_{\gamma}$). The upper-boundary of this region is indicated by the dashed blue line in Figure \ref{fig: ALP-IFMR Constraint}, using values for $T$, $\rho$, $k_s$ and $\kappa_{\gamma}$ taken from our models. Though this does not include any part of the cosmological triangle, it would be interesting to examine how this could be improved upon through detailed stellar simulations.

\subsection{Stellar Rotation}
\label{subsec: rotn}

We have carefully selected the constraint \cite{Andrews} to minimise dependence on astrophysical uncertainties such as star cluster age determination and the initial mass function. Nevertheless, as our investigative tool is stellar modelling, there are systematic uncertainties which we now discuss.

The AGB phase is notoriously difficult to model accurately. The TP-AGB particularly presents significant challenges, owing to its dependence on a combination of intricate processes such as mass loss and enhanced mixing from core overshoot and rotation. An extensive discussion of these can be found in \cite{kerschbaum2007galaxies}.

Typically the uncertainty surrounding free parameters in models of the AGB phase is mitigated by comparison with observation or a solar calibration. Examples of model input physics which fall into this category include the efficiency of mass loss rates and core overshoot. Although their influence on the IFMR can be considerable, observation limits these free parameters to a thin range. Consequentially a discussion of their relevance is deferred to Appendix \ref{sec: AppC}.

Unlike these, however, rotation - and the enhanced mixing it elicits - is a reality of stellar physics. Rather than taking one simple value, the angular velocity of stars will vary according to a probability distribution (e.g. that of \cite{Huang2010}). We therefore present a discussion of the importance of rotational mixing to the IFMR.

\subsubsection*{Effect of rotation on theoretical models}
For simplicity, it has been assumed that all stars simulated in this work are 1-D and do not rotate. The impact of rotation can affect both theoretical predictions for the IFMR and the derivation of its constraints \cite{Cummings_2019}. Rotational mixing increases the supply of hydrogen available when the WD progenitor is evolving on the Main-Sequence, which produces larger stellar cores. This, along with an associated increase in the duration of the E-AGB, means that progenitor stars experiencing rapid rotation will have a higher WD masses than their more slowly rotating counterparts \cite{Dominguez_1996}. Consequently, theoretical predictions of the IFMR including rotation lie above those where it is neglected.

\begin{figure}[t]
    \centering
    \includegraphics{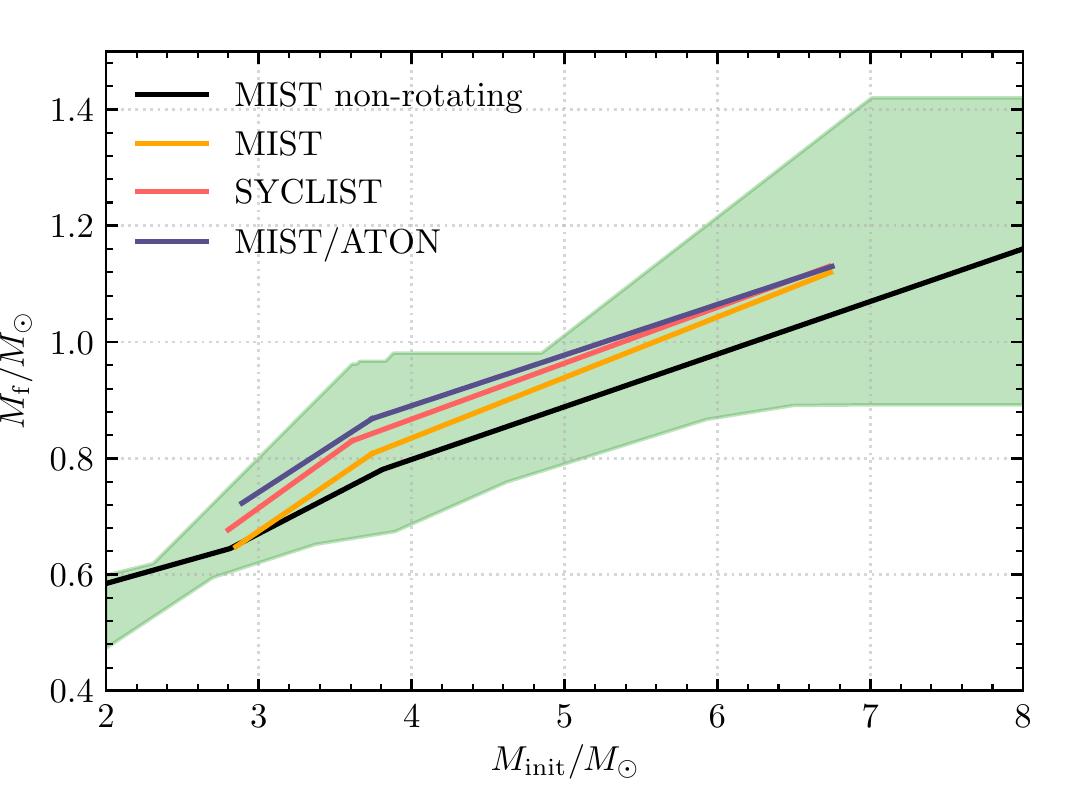}
    \caption{Two-piece fits from the statistical IFMRs of \cite{Cummings_2019} generated using the MIST (yellow), SYCLIST (pink) and MIST/ATON (dark blue) rotating stellar evolution models. The full range of the double white dwarf IFMR constraint \cite{Andrews} and our standard theoretical IFMR (generated from MIST input physics) are also included in green and black respectively. Individual stellar models are referenced in the text.}
    \label{fig: Rotn IFMRs}
\end{figure}

The effects of rotation on theoretical IFMRs was recently investigated in \cite{Cummings_2019}. 40,000 synthetic stars were generated spaced uniformly in initial mass, with rotation drawn from the distribution of \cite{Huang2010}. Values of $M_{\mathrm{f}}$ were then determined through application of three different rotating stellar models - MIST, SYCLIST \cite{SYCLIST1, SYCLIST2} and a combination of ATON/MIST. The two-piece fits of the resulting statistical IFMRs are shown in Figure \ref{fig: Rotn IFMRs}.

The MIST tracks have relatively inefficient rotational mixing, selected to reproduce surface nitrogen abundances in Main Sequence stars \cite{MIST1}. Consequently, the statistical IFMR determined in \cite{Cummings_2019} is weighted strongly towards lower final masses and the two-piece fit in Figure \ref{fig: Rotn IFMRs} is offset from the non-rotating MIST model by an average of $0.04M_{\odot}$ for initial masses between $3.6M_{\odot}$ and $6M_{\odot}$.

The SYCLIST rotating models, however, employ more efficient rotational mixing than those of MIST. Consequently the resulting statistical IFMR is weighted heavily towards larger values of $M_{\mathrm{f}}$. This corresponds to a mean upward shift of $0.06M_{\odot}$ in the same initial mass range. The most significant shift of $0.08M_{\odot}$ is achieved when MIST rotation is applied to ATON models \cite{ATON} and has contributions from both rotational mixing and enhanced convective core overshoot, which we describe in detail in Appendix \ref{sec: AppC}.

In an analysis which accounts for the effects of rotation, the standard astrophysical IFMR is shifted upward from its non-rotating counterpart. Consequently, for theoretical IFMRs to fall outside the region of the DWD constraint, larger values of $g_{a\gamma\gamma}$ are needed than those derived in Section \ref{sec: ALPs and the IFMR}. This could be achieved by repeating our simulations for different stellar rotations and performing an analysis in the spirit of \cite{Cummings_2019}. This approach would be computationally cumbersome and would risk underestimating the influence of rotation owing to the inefficiency of rotational mixing in the MIST input physics. Consequently, we adopt a much more simplistic approach and simply apply an upward shift to our theoretical IFMRs.

We select the magnitude of this upward shift to be $0.08M_{\odot}$, equal to that of the ATON/MIST models and record the new value of $g_{a\gamma\gamma}$ for specific ALP masses which cause the theoretical IFMR to fall outside the DWD constraint. Strictly speaking, the adopted value of $0.08M_{\odot}$ has a contribution of approximately $0.035M_{\odot}$ from the enhanced convective core overshoot of the ATON models. As shall be discussed in Appendix \ref{sec: AppC}, however, the treatment of overshoot in the MIST models has been calibrated to observation during evolutionary phases which are unaffected by ALPs. We are therefore confident that this choice is a conservative one.

\begin{figure}[t]
    \centering
    \includegraphics{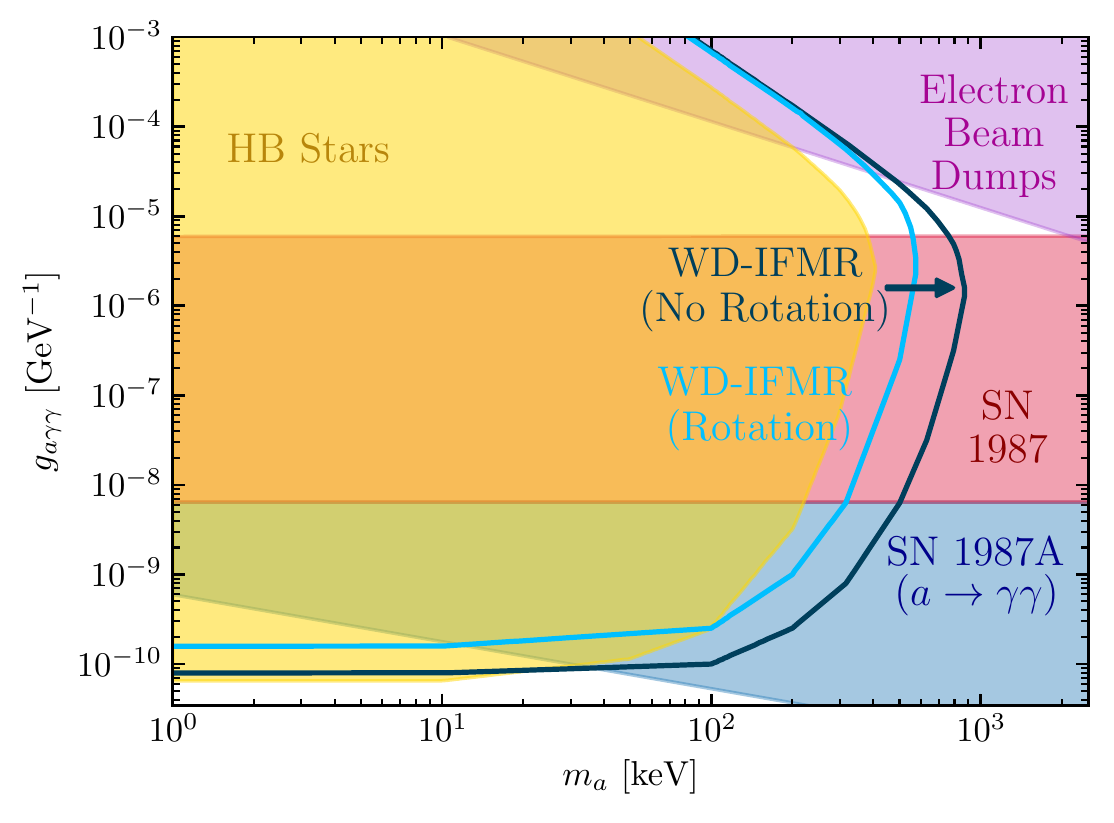}
    \caption{A comparison between our derived constraints including (light blue) and excluding (dark blue) a conservative estimate of the effects of efficient rotational mixing on the IFMR in the keV-MeV region of the ALP-plane.}
    \label{fig: Rotn constraint}
\end{figure}

The updated constraint, which takes these effects into account, is indicated by the light blue line in Figure \ref{fig: Rotn constraint}. While less restrictive than its non-rotating counterpart (indicated by the dark blue line), especially for low ALP masses, we have nevertheless ruled out a significant region of the cosmological triangle.

\subsubsection*{Effect of rotation on the semi-empirical bound}
Stars which rotate have extended progenitor lifetimes due to the enhanced mixing and gravitational lifting effects they experience during their Main Sequence evolution. The faster the rate of rotation, the more extreme this effect. The semi-empirical constraint \cite{Andrews} does not take rotation into account in the \texttt{MESA} models used in their analysis. Given stars in this initial mass range have been observed to rotate with angular velocities of between $\Omega/\Omega_{\mathrm{crit}}=0.0$ and $\Omega/\Omega_{\mathrm{crit}}=0.95$, where $\Omega_{\mathrm{crit}}$ is the critical or break-up angular velocity, it is possible that progenitor lifetimes have been systematically underestimated. As discussed in Section \ref{subsec: comparing theory and observation}, effects which increase progenitor lifetimes shift members of the posterior distribution of \cite{Andrews} downwards, potentially making our bound less restrictive. Consequently we must discuss the expected effects of rotation on the semi-empirical constraint we have employed.

We can estimate the impact of neglecting rotation using the method outlined in Section \ref{subsec: comparing theory and observation}. Stellar models were simulated incorporating rotation of $\Omega/\Omega_{\mathrm{crit}}\in\{0.0,\ 0.2,\ 0.4,\ 0.6,\ 0.8,\\ 0.95\}$. The expected increase in progenitor lifetime was then computed by averaging these over the rotational distributions of \cite{Huang2010}. We found that $\tau_{\mathrm{new}}/\tau$ grows linearly from approximately 1.02 for $2M_{\odot}$ stars to 1.09 for $8M_{\odot}$. From Equation \ref{eq: Mnew}, we find that the corresponding values of $M_{\mathrm{new}}/M$ vary from 1.01 (a 1\% increase) to 1.04 (a 4\% increase) for stars with initial masses between $2$-$8M_{\odot}$. By applying these shifts in initial mass to the posterior sample of IFMRs with fixed breakpoints, we find average downward IFMR shifts of $0.004M_{\odot}$, and $0.017M_{\odot}$ for stars in the $2$-$4M_{\odot}$, $4$-$8M_{\odot}$ initial mass ranges respectively. Of these, it is the latter that is the most important for our constraint.

Given ALPs shift the IFMR downwards, it is more pertinent to discuss the impact of rotation on the lower boundary of the constraint \cite{Andrews}. We achieve this by estimating the effect of this increase in progenitor lifetimes on the lower 95\% confidence interval boundary of the posterior distribution (see Figure \ref{fig: IFMR prob} in Appendix \ref{sec: AppD}). In the $2$-$4M_{\odot}$ and $4$-$8M_{\odot}$ initial mass ranges, we find these values reduce by an average of $0.003M_{\odot}$ and $0.013M_{\odot}$ respectively. For the most massive stars, this boundary has a smaller downward shift than that of the average over the posterior distribution. This is because IFMRs towards the lower end of the distribution are generally flatter, and therefore must shift less vertically to accommodate the increase in $M_{\mathrm{init}}$ described above.

Stellar rotation is clearly a significant source of systematic uncertainty in our analysis. The results above, however, suggest that it has a greater influence on the predictions of our simulations rather than the derivation of semi-empirical constraints. In principle, a comprehensive analysis should account for the latter directly. However, given the magnitude of these effects is relatively small, our original analysis excluded the entire range of IFMRs with unfixed breakpoints (rather than a confidence interval) and we have already opted for an upward shift of $0.08M_{\odot}$ in this section (which accounts for the combined effects of rotation and enhanced core overshoot), we ignore this contribution in our final result.


We stress that a conservative approach has been adopted throughout this analysis. We chose to use the DWD binary constraint \cite{Andrews} over others as it is insensitive to star cluster ages and has unfixed breakpoints which leads to a less restrictive range of IFMRs. We have also allowed for the possibility of efficient rotational mixing affecting our predictions. Naturally, there is scope for this constraint to improve dramatically when theoretical uncertainties surrounding AGB physics and rotational mixing are better understood. Furthermore, millions of binary systems have been resolved in the Gaia data release 3 \cite{2021arXiv210105282E}. Given this analysis was performed with only 14 binary systems, it would be of great interest for the method of \cite{Andrews} to be applied to the wide subset of the 1400 double white dwarf binaries identified.

%% file: Section5/Sec5.tex
\section{Beyond the Initial-Final Mass Relation}
\label{sec: Section 5}

In Section \ref{sec: Section 4} we elected to use the white dwarf initial-final mass relation as the basis of our constraint. However, the behaviour discussed in Section \ref{sec: Sec 3} has other potential observable effects. Two of these, which were proposed in \cite{Dominguez:2017mia}, reference the impact ALPs can have on the ultimate fates of stars.

The final state into which a star evolves depends sensitively on the nature of carbon fusion, which is principally governed by the CO core mass. Stars with $M_{\mathrm{CO}}<1.06M_{\odot}$ never reach the requisite conditions for carbon ignition and end their lives as CO white dwarfs. Alternatively, if $1.06M_{\odot}<M_{\mathrm{CO}}<1.38M_{\odot}$, the CO core becomes partially degenerate and carbon ignition occurs in an off-centre flash. The inner boundary of this burning region then advances to the centre of the star, paving the way for stable carbon burning and the development of an oxygen-neon (O-Ne) core. Such stars are termed \textit{Super-AGB} stars and are the progenitors for O-Ne white dwarfs. If the CO core is still more massive after the exhaustion of central helium ($M_{\mathrm{CO}}>1.38M_{\odot}$), carbon is ignited in a stable, convective core. Such stars experience all further episodes of nuclear burning and are the progenitors of core collapse supernovae (CCSN)\footnote{Not all stars in this mass range do experience a supernova (see e.g. \cite{2009ARA&A..47...63S}), however this is recognised as the minimum CO core mass for such an event.}.

A great deal of work has been conducted in astrophysics to identify the masses, $M_{\mathrm{up}}$ and $M_{\mathrm{up}}'$, which correspond to the minimum mass of O-Ne white dwarf and CCSN progenitors respectively (for a review see \cite{2009ARA&A..47...63S}). These values naturally vary between models, though it is believed that they lie between $7$-$8M_{\odot}$ and $10$-$12M_{\odot}$ respectively \cite{2012sse..book.....K}.

As the predicted value of $M_{\mathrm{CO}}$ varies significantly when ALPs are included in the stellar model (see Section \ref{sec: Sec 3}), so too do the values of $M_{\mathrm{up}}$ and $M_{\mathrm{up}}'$. Specifically, given that ALP-production reduces the mass of $M_{\mathrm{CO}}$, much larger initial masses are required for the formation of O-Ne WD and CCSN progenitors. This impact, in the context of ALPs below the keV-MeV scale, is the subject of \cite{Dominguez:2017mia}, wherein it is suggested that observational upper limits on $M_{\mathrm{up}}'$, or the rates of Type Ia SN could be used to constrain ALPs. Here we shall briefly comment on the prospects and potential pitfalls of constructing a constraint on ALPs based on these and two other pieces of observational evidence.

\subsubsection*{Core Collapse Supernova Progenitors:}
Theoretical values of $M_{\mathrm{up}}'$, the minimum mass of CCSN progenitors, vary between $10$-$12M_{\odot}$. As such stars reach temperatures substantially higher than in the He-B shells of AGB stars, constructing a constraint based on heavy ALP production in their interiors is an encouraging prospect. However, whether or not such observational constraints can be used depends on the method of their construction. There appear to be two sources of such constraints used in the literature.

The first of these is via the investigation of pre-explosion images. If a CCSN is detected in a region which has previously been photographed, the progenitor candidate can be analysed and an initial mass estimated through a theoretical initial mass-final luminosity relation. In \cite{Smartt2009} this led to a constraint of $8^{+1.0}_{-1.5}M_{\odot}$. The initial mass-final luminosity relation, however, relies on the predictions of stellar evolution models during the AGB and have already been shown to vary when ALP-production is included in the simulations \cite{Straniero:2019dtm}. To prevent self-consistency issues, the analysis in \cite{Smartt2009} would have to be re-derived using an initial mass-final luminosity relation which takes the ALP-dependence into account.

Another source of constraints on $M_{\mathrm{up}}'$ comes from the analysis of supernova remnants (SNRs), e.g. \cite{Diaz-Rodriguez2018-SNRs} which finds $M_{\mathrm{up}}'=7.33^{+0.02}_{-0.16}M_{\odot}$. Key components used in the derivation of this limit are star formation histories (SFHs), which require the use of stellar isochrones. As these isochrones are themselves derived from stellar modelling, it would be necessary to ascertain the degree to which these depend on ALP properties before this constraint could be used. As a minimal requirement, consistency demands that the SFHs be determined using isochrones derived from the same stellar evolution code used to analyse the impact of ALPs (in our case the MIST isochrones). This is further complicated by the dependence of $M_{\mathrm{up}}$ and $M_{\mathrm{up}}'$ on the $^{12}$C+$^{12}$C reaction rate, which is presently a source of great uncertainty in stellar simulations \cite{inbook}.

\begin{figure}[t]
    \centering
    \includegraphics{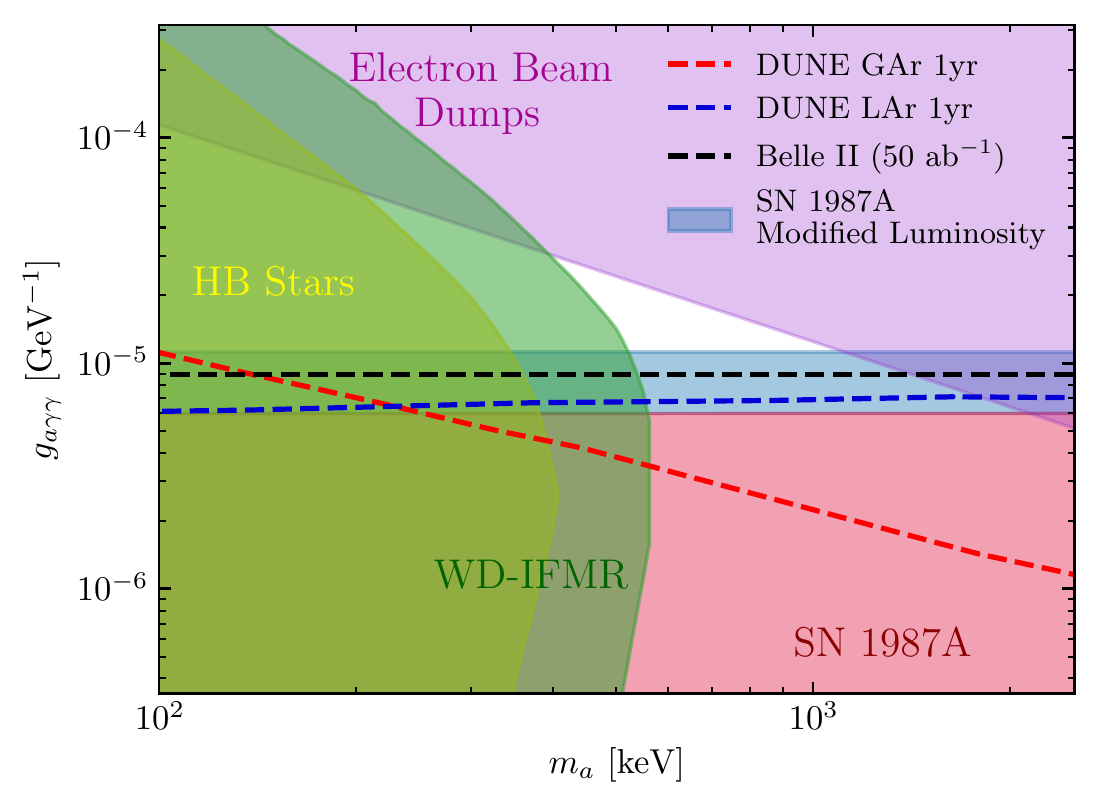}
    \caption{The state of the ALP plane in the keV-MeV region including our constraint (Section \ref{sec: Section 4}), the modified luminosity SN1987A constraint (shaded purple) \cite{Lucente:2020whw}. The projected sensitivity of Belle II \cite{Dolan:2017osp}, and DUNE liquid (LAr) and gaseous argon (GAr) detectors \cite{Brdar:2020dpr} are indicated by the region above the dashed black, blue and red lines respectively.}
    \label{fig: ALP-Decay Constraint}
\end{figure}

\subsubsection*{Type IA Supernova Rates:}
Type Ia SN are believed to occur when the mass of a CO white dwarf in a binary system exceeds the Chandrasekhar limit due to accretion from a main sequence or giant partner (single degenerate pathway) or a merger with a second white dwarf (double degenerate pathway). For a detailed review of these pathways see \cite{WANG2012122}. 

Given the inclusion of ALPs in stellar models leads to larger $M_{\mathrm{up}}$ values, their existence would favour greater populations of CO white dwarfs and, consequently, a higher incidence of Type Ia SN. It is therefore conceivable that bounds on the latter could enable the properties of ALPs to be constrained or, as suggested in \cite{Dominguez:2017mia}, might even hint to their existence.

Constraints on Type Ia SN are typically presented in terms of their delay time distribution, i.e. their rate of incidence as a function of time following a normalised burst of star formation (see \cite{doi:10.1146/annurev-astro-082812-141031} for a more detailed description). It would be interesting to explore the manner in which ALP-production affects the predicted shape of the delay time distribution. Such an analysis would, naturally, require ALPs to be included in binary stellar modelling. A complete analysis on the topic would naturally have to investigate both production pathways.

\subsubsection*{Mira Variable Drifting:}
Mira variable stars are a sub-class of asymptotic giants which are unstable to radial pulsations with periods that are $\mathcal{O}(100\ \mathrm{days})$. Certain Mira variables have descriptive data which go back over a century. As a result, it has been possible to detect period-drifting in these stars, the magnitude of which can be stark. For instance, the period of R Hya has changed from approximately 500 days, as measured in 1700, to 387 at the year 2000 \cite{AGBStarsBook}. Similarly, R Aql's period has decreased from approximately 320 to 267 days in recordings between 1915 and 2000.

It was shown in \cite{WoodMiras1981} that this drifting is consistent with helium shell flashes occurring during the larger-scale thermal pulsations of the TP-AGB, though this is still an open area of debate in the literature \cite{NeilsonMiraPeriodChange}. It has previously been shown \cite{Dominguez}, and we confirmed in Section \ref{sec: Sec 3}, that the thermal pulses of stars with a given initial mass vary when ALPs are included in the model. It would be interesting to investigate whether these changes introduce tension between the agreement of period-drifting and long-term evolution. This likely requires the inclusion of ALP-production within dedicated TP-AGB models.

\subsubsection*{Abundance Ratios:}
A further consequence of the inclusion of ALP-production in stellar modelling is the impact it has on elemental abundance ratios within the star. By accelerating periods of helium burning (both central and shell), the relative abundance of carbon and oxygen in the core is likely to change. A deeper dredge-up event is also likely to increase the presence of fusion products in the surface composition. In fact, the presence of light ALPs in the late-evolutionary phases of $16M_{\odot}$ has already been found to dramatically increase the abundance of oxygen, magnesium and neon for values of $g_{a\gamma\gamma}$ as low as $10^{-11}~\mathrm{GeV}^{-1}$ \cite{Aoyama:2015asa} in the case of the latter. This is likely exacerbated by ALP-enhanced second and third dredge-up events which increase the presence of fusion products at the stellar photosphere.

Observational measurements of these ratios could therefore constitute a potent source of constraints on ALP parameter space. Unfortunately information about AGB stars themselves is somewhat scarce, with examples limited to individual post-AGB stars, e.g. \cite{PostAGBSpec}. This could be mitigated through comparison of theory with abundance constraints from spectroscopic analysis of white dwarfs such as \cite{Coutu_2019} or via the investigation of the composition of planetary nebula (see for example \cite{PlanNebAGB, karakas_lattanzio_2003}).

It is likely, owing to the sensitivity of dredge-up events to the adopted prescription of core overshoot (see Appendix \ref{sec: AppC}) that this would introduce significant systematic uncertainty. Despite this, the use of spectroscopic analysis is a particularly compelling prospect, as it does not rely on any of the results of stellar evolution theory, circumventing any issues of self-consistency.

%% file: Section6/Sec6.tex
\section{Conclusion}
\label{sec: Section 6}

Stellar evolution has a well-established pedigree in constraining physics beyond the Standard Model, most notably for axions and axion-like particles. In this work we provide a detailed investigation of the effects of keV-MeV scale ALPs on stellar evolution simulations.

The photo-production of such axion-like particles in the keV-MeV mass range significantly impacts the evolution of asymptotic giant branch stars, the late evolutionary phase of stars with initial masses $\lesssim8M_{\odot}$. Specifically, the free streaming of ALPs produced in the helium-burning shells of these stars facilitates more rapid and deeper dredge-up events, which significantly reduce their final masses.

This behaviour has been constrained by appealing to semi-empirical measurements of the white dwarf initial-final mass relation. In particular, analysis of 14 wide double white dwarf binary systems conducted in \cite{Andrews} enabled us to construct a new bound on the ALP-plane which proves more restrictive for large $m_a$ than that derived from horizontal branch stars, most notably in the unconstrained cosmological triangle, even when the effects of stellar rotation have been considered. We expect these results to improve in the near future if the method of \cite{Andrews} were applied to a subset of the 1400 double white dwarf binaries identified in the Gaia early Data Release 3 \cite{2021arXiv210105282E}.

For sufficiently large values of $g_{a\gamma\gamma}$ and $m_a$, the axion-like particle decay-lengths fall below the width of the helium-burning layer and the foundational criterion of the energy-loss argument is no longer met. This reduces our initial constraint to the green shaded region in Figure \ref{fig: ALP_param_space}. As energy transfer within the helium-burning layer is radiative, more strongly interacting axion-like particles still influence the structural evolution of asymptotic giant branch stars. We can estimate the region of the ALP-plane in which this is relevant by insisting that the Rooseland mean opacity of axion-like particles be larger than that of photons. A conclusive statement about these effects would require the addition of axion-like particle energy transfer to stellar models.

Within the last year there has been a resurgent interest in the cosmological triangle. In addition to the recent HB star bound \cite{Carenza:2020zil}, which our work complements, the constraints which define its other boundaries have been revisited. For instance, the constraint derived from the neutrino signal and observed cooling of SN1987A was recomputed recently with a state-of-the-art supernova model \cite{Lucente:2020whw}, which we show in Figures \ref{fig: ALP_param_space}, \ref{fig: ALP-IFMR Constraint} and \ref{fig: ALP-Decay Constraint}. The work in question included a second calculation, based on the condition that only the part of the axion-like particle luminosity that can be readily converted to neutrino energy is relevant for the SN1987A bound \cite{Chang:2016ntp, Ertas:2020xcc}. When this so-called \textit{modified luminosity} criterion is applied a new region above the pre-existing constraint is excluded, and the cosmological triangle shrinks further (see Figure \ref{fig: ALP-Decay Constraint}).

The definition of new boundaries for the cosmological triangle is timely, as future experiments will be able to directly probe this region. The Belle II experiment, for example, has estimated sensitivity within the relevant mass range to couplings as low as $g_{a\gamma\gamma}\sim10^{-5}$ GeV$^{-1}$ at a luminosity of 50 ab$^{-1}$ \cite{Dolan:2017osp}. Future neutrino experiments, such as DUNE, will also be able to probe the cosmological triangle with both liquid argon (LAr) and gaseous argon (GAr) detectors \cite{Brdar:2020dpr}.

This work, like many before it, employs robust observational astrophysics and ever-more accessible stellar modelling to investigate the impact of axion-like particles on stellar evolution. Though we have directed these tools towards this class of particle, the impact of other weakly interacting particles can be probed in this manner. Stars, as ubiquitous objects in the universe, have a vital role to play in constraining physics beyond the Standard Model.

%% file: AppA/AppA.tex
\section{Details of Simulations}

\subsection{Treatment of energy-loss in \texttt{MESA}}
\label{Sec: AppA}
Our treatment of energy-loss to ALP-production within \texttt{MESA} followed the example set by \cite{Friedland:2012hj}, which augmented the module responsible for thermal neutrino rates (\texttt{neu}) and added an additional term corresponding to $\epsilon_a$. The rationale behind this was simply that power loss to neutrinos and freely escaping ALPs yield virtually identical phenomenological effects on stellar evolution, in that they both contribute a negative term to the total energy production rate $\epsilon$.

The functions $F(\xi^2, \mu^2)$ and $G(\mu^2)$, defined in Equations \ref{eq: F function} and \ref{eq: G function} respectively, contain the entire chemical composition and ALP mass dependence of $\epsilon_a$. Recall that $\xi=k_s/(2T)$ and $\mu=m_a/T$. We took the sum of these functions
\begin{equation}
    H(\xi^2, \mu^2)=F(\xi^2, \mu^2)+G(\mu^2) 
\end{equation}
and evaluated it at fixed points, logarithmically spread in $\xi$ and $\mu$, to define a grid which is loaded the first time the \texttt{neu} routine is called. Within each cell of the \texttt{MESA} model, our augmented \texttt{neu} routine determines the appropriate values of $\xi$ and $\mu$ (cell average properties) and interpolates between our grid points to find the corresponding value of $H(\xi^2, \mu^2)$ and hence $\epsilon_a$. This is added to the energy-loss rate to thermal neutrinos, as determined by the standard \texttt{neu} routine.

The aforementioned interpolation is carried out using the PSPLINE\footnote{\url{https://w3.pppl.gov/\~pshare/help/pspline.htm}} bicubic spline algorithm in \texttt{MESA}'s own \texttt{interp\_2d} routine (see \cite{MESA1} for more details). We compared the $\epsilon_a$ values in several of our models to its numerically evaluated counterpart and found it matched to within 0.2\%.
 
As per the \texttt{MESA} manifesto, we have made our \texttt{run\_star\_extras} file, which contains the modified \texttt{neu} routine, as well as our grid of pre-calculated $H(\xi^2, \mu^2)$ values available for download\footnote{\url{https://github.com/fhiskens/MESA\_ALPs}}.

\subsection{Adopted input physics}
\label{sec: AppB}

In addition to making our extension to MESA publicly available, we also include the \texttt{inlist} containing our input physics, which has been taken from the MIST project. The details of these choices as well as their motivation can be found in \cite{MIST0,MIST1}. As all simulations have initial masses below $10M_{\odot}$, we choose the prescription intended for low and intermediate mass stars. Aspects of our \texttt{run\_star\_extras} file also originated with MIST, particularly those which assist with conversion of the stellar model and the switching of boundary conditions after 100 steps of the simulation.

\texttt{MESA} was compiled using the \texttt{MESA} SDK\footnote{\url{http://www.astro.wisc.edu/~townsend/static.php?ref=mesasdk}} \cite{richard_townsend_2019_2669541}. All results were analysed using the \texttt{mesa\_reader} Python package\footnote{\url{https://github.com/wmwolf/py\_mesa\_reader}}.

%% file: AppB/AppB.tex
\section{Systematic Uncertainties}
\label{sec: AppC}
In Section \ref{subsec: rotn} we accounted for the influence of rotation on our constraint. However, there are multiple free parameters in our models which can influence the IFMR. Here we discuss two of these - mass loss and convective overshoot - and estimate their influence on our constraint.

\subsubsection*{Mass loss:}
The MIST models \cite{MIST0, MIST1} on which we base our input physics adopt the Reimers \cite{1975MSRSL...8..369R} and Bl\"{o}cker \cite{1995A&A...297..727B} prescriptions for mass loss for the RGB and AGB respectively. These are given by
\begin{equation}
    \Dot{M}_{\mathrm{R}}=4\times10^{-13}\eta_{\mathrm{R}}\frac{(L/L_{\odot})(R/R_{\odot})}{(M/M_{\odot})}\ \mathrm{M}_{\odot}\ \mathrm{yr}^{-1}
    \label{eq: Reimers}
\end{equation}
and
\begin{equation}
    \Dot{M}_{\mathrm{B}}=4.83\times10^{-9}\eta_{\mathrm{B}}\frac{(L/L_{\odot})^{2.7}}{(M/M_{\odot})^{2.1}}\frac{\Dot{M}_{\mathrm{R}}}{\eta_{\mathrm{R}}} \mathrm{M}_{\odot}\ \mathrm{yr}^{-1}
    \label{eq: Blocker}
\end{equation}
where $\eta_{\mathrm{R}}$ and $\eta_{\mathrm{B}}$ are $\mathcal{O}(1)$ parameters. In the MIST models, values of $\eta_{\mathrm{R}}=0.1$ and $\eta_{\mathrm{B}}=0.2$ have been chosen to reproduce the IFMR and AGB luminosities in the Magellanic Clouds.

Mass loss rates are relevant to the IFMR, as they govern how quickly the outer envelope is shed during the TP-AGB and consequently how many thermal pulses occur during this phase. Indeed, separating the effects of mass-loss and ALPs was viewed as the major challenge in establishing an IFMR-derived constraint on ALP parameters when discussed in \cite{Dominguez} for low mass ALPs.

Varying the magnitude of  $\eta_{\mathrm{B}}$ does impact the lower-IFMR ($M_{\mathrm{init}}\lesssim3M_{\odot}$), with more efficient mass loss leading to lower WD masses for a given value of $M_{\mathrm{init}}$ \cite{Cummings_2019}. For larger initial masses, however, the IFMR remains relatively insensitive to the adopted mass loss rate as the TP-AGB of such stars is too rapid for significant core growth to occur \cite{1995A&A...297..727B}.

\begin{figure}[t]
    \centering
    \includegraphics{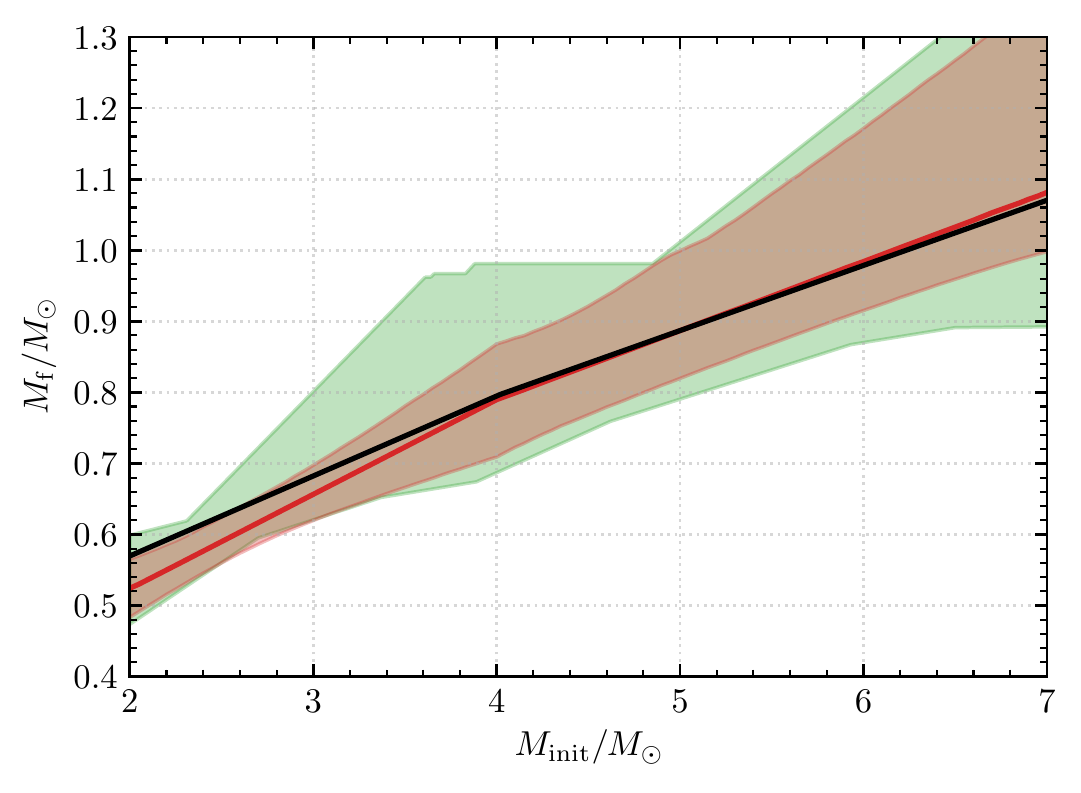}
    \caption{A comparison between double white dwarf constraints from \cite{Andrews}. The full range of the bound where breakpoints are allowed to move is shown in green. The 95\% confidence interval for fixed breakpoints at $2M_{\odot}$ and $4M_{\odot}$ is shown in red. Its average is indicated by the red line. The black solid line is a two-piece linear fit derived from our simulations (the black points in Figure~\ref{fig: IFMR fit}), constrained with a breakpoint fixed at $4M_{\odot}$. }
    \label{fig: IFMR prob}
\end{figure}

\subsubsection*{Convective overshoot:}
Fluid parcels in a convective region are accelerated as they approach its boundary and do not begin to decelerate until they enter the radiative zone. Consequently, if braking is insufficient, they can penetrate a non-negligible distance beyond the convective boundary and increase the efficiency of mixing in this region. This phenomenon is known as convective overshoot \cite{2012sse..book.....K}.

The presence of convective overshoot has two main results on the IFMR. Firstly, the enhanced mixing caused by overshoot facilitates the formation of more massive CO core at the onset of the TP-AGB, which directly shifts the IFMR upwards \cite{2000A&A...360..952H}. On the other hand, increased overshoot during the TP-AGB causes deeper third dredge-up events, which reduces core growth during this phase \cite{2000A&A...360..952H, Cummings_2019}.

The MIST models treat overshoot as a time-dependent, diffusive process, the strength of which is governed by a parameter $f_{\mathrm{ov}}$. For the core, envelope and shell, values of $f_{\mathrm{ov, core}}=0.016$ and $f_{\mathrm{ov, env}}=f_{\mathrm{ov, shell}}=0.0174$ respectively. These are adopted in order to reproduce the Main Sequence turn-off of open cluster Messier 67 ($f_{\mathrm{ov, core}}$) and a solar calibration ($f_{\mathrm{ov, env}}$) \cite{MIST1}.

The ATON models of \cite{ATON} similarly model overshoot as a diffusive process. However, these have $f_{\mathrm{ov}}=0.02$ for hydrogen and helium burning and $0.002$ during the AGB. As a result, they predict an IFMR which is higher than that of the MIST models by an average of $0.035M_{\odot}$ for initial masses between $3.6$-$6.5M_{\odot}$. This contributes to the mean upward shift of $0.08M_{\odot}$ corresponding the combined MIST/ATON models, which was chosen for our constraint in Section \ref{subsec: rotn}.

Enhanced convective core overshoot can also affect progenitor lifetimes relevant to the derivation of semi-empirical IFMR constraints. By essentially enhancing the size of the convective core, a larger supply of hydrogen fuel is available for Main Sequence stars, which increases the duration of this evolutionary phase. To estimate the magnitude of this effect, we recomputed our evolutionary models adopting the values of $f_{\mathrm{ov}}$ specified in the ATON models. The resulting stellar lifetimes were found to be between 1-2\% larger for stars in the $2$-$8M_{\odot}$ initial mass range. Following the argument presented in Section \ref{subsec: comparing theory and observation}, this corresponds to $M_{\mathrm{new}}/M_{\mathrm{init}}$ values increasing linearly from 1.01 to 1.03 for the same range of initial masses. Again, we estimate the influence this has on the constraint of \cite{Andrews} by averaging the resulting downward shift in $M_{\mathrm{WD}}$ over the entire posterior sample of IFMRs with fixed breakpoints. For stars between $2$-$4M_{\odot}$ and $4$-$8M_{\odot}$ we find average downward shifts of $0.001M_{\odot}$ and $0.004M_{\odot}$ respectively. These are an order of magnitude lower than the upward shift adopted in our constraint, and are less than the upward shift obtained when the effects of ALPs on progenitor lifetimes are considered. 

Importantly, both mechanisms for tuning the parameter $f_{\mathrm{ov}}$ pertain to the Main Sequence. They are therefore insensitive to ALPs in the keV-MeV mass range, the production of which is Boltzmann suppressed during evolutionary phases before the AGB. Even ALPs lighter than this, which are readily produced during central helium-burning, will minimally affect the Main Sequence and therefore will not invalidate these calibrations. Because of this, we are confident that the adopted upward shift of $0.08M_{\odot}$ used in Section \ref{subsec: rotn} is conservative.

%% file: AppC/AppC.tex
\section{A Probabilistic Approach}
\label{sec: AppD}

\begin{figure}[t]
    \centering
    \includegraphics{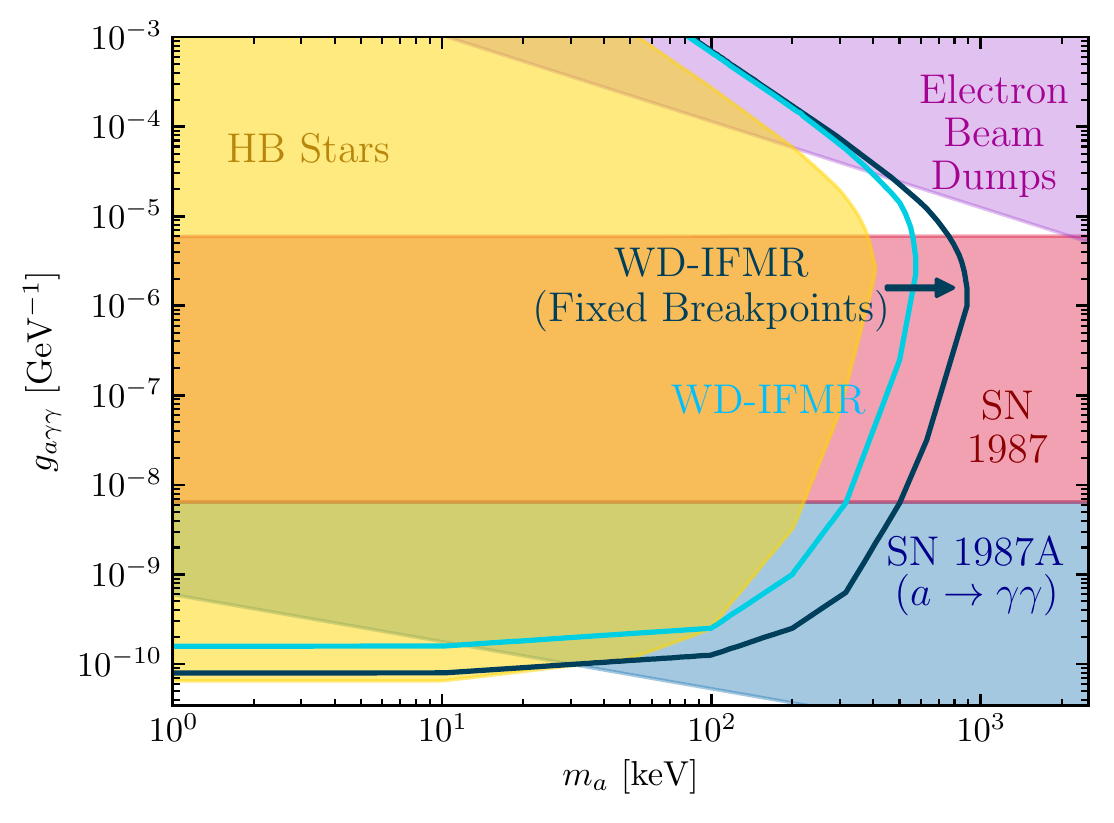}
    \caption{A comparison between our derived constraint from \ref{subsec: rotn} (light blue) and that when a probabilistic approach is adopted (dark blue) in the keV-MeV region of the ALP-plane. Both constraints conservatively account for rotational mixing.}
    \label{fig: Prob constraint}
\end{figure}

In Section \ref{sec: Section 4} we adopted the less restrictive IFMR constraint from \cite{Andrews} in which the breakpoints are allowed to vary. Given the posterior sample pertaining to this constraint was not available, instead we conservatively insisted that theoretical IFMRs must fall within its entire range. The posterior sample with fixed breakpoints at $2M_{\odot}$ and $4M_{\odot}$, however, was made available, which enables us to establish what a probabilistic approach would entail.

We first define a 95\% confidence interval about the mean for uniformly spaced values of $M_{\mathrm{init}}$. This is shown by the red region in Figure \ref{fig: IFMR prob}, with mean indicated by the solid red line. The region corresponding to unfixed breakpoints is also included in green. The black solid line is a two-piece fit derived from our simulations (the black points in Figure \ref{fig: IFMR fit}), constrained with a breakpoint fixed at $4M_{\odot}$.

We can then repeat the analysis detailed in Section \ref{sec: ALPs and the IFMR} for a two-piece fit with breakpoint at $4M_{\odot}$ in order to generate a new result. The corresponding constraint, adjusted for ALP-decay and the influence of efficient rotation, is shown in Figure \ref{fig: Prob constraint} in dark blue. Also included is the constraint from Section \ref{subsec: rotn} in light blue.

By adopting this probabilistic approach, our constraint becomes considerably more restrictive. While the HB star bound is still more restrictive at low mass, it is worth noting that we have accounted for efficient rotational mixing in this calculation. If the treatment of rotation is relaxed entirely our constraint moves down to $g_{a\gamma\gamma}=0.316\times10^{-10}$ GeV$^{-1}$, below that of the HB star bound $g_{a\gamma\gamma}=0.66\times10^{-10}$ GeV$^{-1}$. A definitive statement on the matter, however, requires a deeper understanding of rotational mixing in stars.

Although we do not adopt this probabilistic approach for our main constraint, it demonstrates clearly the possible constraining power of the IFMR. It should be noted that the constraint \cite{Andrews} could be improved in the near future if it is applied to the larger double white dwarf binary dataset recently identified in the Gaia early Data Release 3 \cite{2021arXiv210105282E}. Such an analysis would be of great interest to particle physicists hoping to constrain axion-like particles.